\documentclass[titlepage,12pt]{article}
\usepackage{amssymb,amsmath,color,graphics,amscd,epsf,indentfirst,amsfonts}
\usepackage{enumerate}
\usepackage{epsfig}
\usepackage{hyperref}
\usepackage{cite}
\usepackage{scalefnt}
\usepackage[titles]{tocloft}
\usepackage{sectsty}
\allsectionsfont{\bf \scalefont{.7} \selectfont}
\subsectionfont{\bf \scalefont{.85} \it \selectfont}
\subsubsectionfont{\bf \scalefont{1} \it \selectfont}
\usepackage[T1]{fontenc}
\usepackage{lmodern}
\usepackage{bibspacing}

\def\blfootnote{\xdef\@thefnmark{}\@footnotetext}

\long\def\symbolfootnote[#1]#2{\begingroup%
\def\thefootnote{\fnsymbol{footnote}}\footnote[#1]{#2}\endgroup}

\makeatletter
\renewcommand{\@dotsep}{4.5}
\makeatother
\def\be{\begin{equation}}
\def\ee{\end{equation}}

\makeatletter
\def\@seccntformat#1{\csname the#1\endcsname.\quad}
\makeatother

\def\clock{{\count0=\time
           \divide\count0 60
           \ifnum\count0<10 0\fi\the\count0
           \multiply\count0 -60 \advance\count0 \time
           :\ifnum\count0<10 0\fi \the\count0
         }}
\newcommand{\timestamp}{{\small\vbox{\hbox{\tt\jobname.tex}
\hbox{\the\day/\the\month/\the\year, \clock}}}}

\def\CC{{\cal C}}

\def\FF{{\cal F}}

\def\HH{{\cal H}}
\def\II{{\cal I}}

\def\MM{{\cal M}}
\def\NN{{\cal N}}
\def\OO{{\cal O}}

\def\SS{{\cal S}}

\def\d{{\partial}}

\newcommand{\beq}{\begin{equation}}
\newcommand{\eeq}{\end{equation}}
\newcommand{\ba}{\begin{array}}
\newcommand{\ea}{\end{array}}
\newcommand{\bea}{\begin{eqnarray}}
\newcommand{\eea}{\end{eqnarray}}

\newcommand{\C}{\mathbb{C}}
\newcommand{\R}{\mathbb{R}}

\newcommand{\N}{\mathbb{N}}

\newcommand{\tr}{\mathop{{\rm Tr}}}

\numberwithin{equation}{section}
\begin{document}

\begin{titlepage}
\begin{flushright}
{CCTP-2011-24}
\end{flushright}
\vskip 3.3cm
\begin{center}
\font\titlerm=cmr10 scaled\magstep4
    \font\titlei=cmmi10 scaled\magstep4
    \font\titleis=cmmi7 scaled\magstep4
    \centerline{\titlerm
     F-theorem, duality and SUSY breaking in}
           \vspace{0.4cm}
     \centerline{\titlerm 
     one-adjoint Chern-Simons-Matter theories}
\vskip 1.5cm
{\it Takeshi Morita and Vasilis Niarchos}\\
\vskip 0.7cm
\medskip
{Crete Centre for Theoretical Physics,}\\ 
{Department of Physics, University of Crete, 71003, Greece}\\
\medskip
{takeshi@physics.uoc.gr, niarchos@physics.uoc.gr}

\end{center}
\vskip .4in
\centerline{\bf Abstract}

\baselineskip 20pt
%
%\date{}

\vskip .5cm \noindent
We extend previous work on $\NN=2$ Chern-Simons theories coupled to a single 
adjoint chiral superfield using localization techniques and the $F$-maximization principle. 
We provide tests of a series of proposed 3D Seiberg dualities and a new class 
of tests of the conjectured F-theorem. In addition, a proposal is made for a modification of the 
$F$-maximization principle that takes into account the effects of decoupling fields. Finally, we 
formulate and provide evidence for a new general non-perturbative constraint on spontaneous 
supersymmetry breaking in three dimensions based on $Q$-deformed $S^3$ partition functions. 
An explicit illustration based on the known analytic solution of the Chern-Simons matrix model is 
presented.

\vfill
\noindent
August, 2011
\end{titlepage}\vfill\eject

\setcounter{equation}{0}

\pagestyle{empty}
\small
\tableofcontents
\normalsize
%\newpage
\pagestyle{plain}
\setcounter{page}{1}

%%%%%%%%%%%%%%%%%%%%%%%%%%%%%%%%
\section{Introduction}
\label{intro}

Important progress has been recently achieved in understanding some of the key properties of 
three-dimensional supersymmetric conformal field theories (SCFTs). The main technical tool has 
been localization which (after proper regularization) allows to express the free energy of 
the Euclidean CFT on the three-sphere
\beq
\label{introaa}
F=-\log |Z_{S^3}|
\eeq
in terms of a compact matrix integral expression \cite{Kapustin:2009kz,Jafferis:2010un,Hama:2010av}.

The free energy $F$ provides a variety of useful information. For example, 
three-dimensional SCFTs with $\NN=2$ supersymmetry have a conserved $U(1)$ R-symmetry that
sits in the same supermultiplet as the stress-energy tensor. In a general interacting SCFT this symmetry
receives quantum corrections and the quantum numbers associated with it become non-trivial 
functions of the parameters that define the theory. The proper coupling of the matter fields to the 
curvature of $S^3$ makes $F$ a function of the R-charges. It has been proposed that the exact 
$U(1)$ R-symmetry of the theory is that R-symmetry which locally maximizes $F$ \cite{Jafferis:2010un}. 
Non-trivial tests of this proposal have appeared in \cite{Martelli:2011qj,Cheon:2011vi,
Jafferis:2011zi,Amariti:2011hw,Niarchos:2011sn,Minwalla:2011ma,Amariti:2011xp}. 

Secondly, there are by now many known examples of three-dimensional SCFTs that exhibit dualities 
analogous to the S-duality and Seiberg duality in four dimensions. Matching the free energy $F$ of 
dual theories provides non-trivial checks of these dualities 
\cite{Kapustin:2010xq,Kapustin:2010mh,Willett:2011gp,Kapustin:2011gh,Benvenuti:2011ga}
(analogous checks of 3D dualities based on the computation of superconformal indices,
see $e.g.$ \cite{Bhattacharya:2008zy}, have been discussed in 
\cite{Hwang:2011qt,Krattenthaler:2011da}). 
 
Moreover, it has been noted that there is a close relation between superconformal
indices in four dimensions \cite{Kinney:2005ej,Romelsberger:2005eg} and sphere partition 
functions in three dimensions \cite{Dolan:2011rp,Gadde:2011ia,Imamura:2011uw}. In this context,
the mathematical identities that relate the superconformal indices of dual theories in four dimensions
are closely related to the mathematical identities that relate the sphere partition functions of dual
theories in three dimensions. This observation raises the possibility of a promising 3D-4D connection.

Finally, it has been proposed \cite{Casini:2011kv,Jafferis:2011zi} that $F$ is a quantity that plays in 
three dimensions the same role that the $c$-function plays in two dimensions.\footnote{The $F$-
theorem was first argued as a general $a_d^*$-theorem in the context of holography in 
\cite{Myers:2010xs,Myers:2010tj}.} In analogy 
to Zamolodchikov's $c$-theorem in two dimensions \cite{Zamolodchikov:1986gt}
(or the $g$-theorem for one-dimensional boundary RG flows \cite{Affleck:1991tk,Affleck:1992ng}), an 
$F$-theorem has been conjectured in three dimensions, which states that $F$ decreases along RG 
flows. Successful checks of this proposal have appeared in 
\cite{Gulotta:2011si,Amariti:2011xp,Klebanov:2011gs}.

In this paper we will discuss all three of the above aspects in the context of a special class of $\NN=2$ 
Chern-Simons-Matter (CSM) theories defined as $\NN=2$ Chern-Simons theory at level $k$ coupled 
to a single chiral superfield in the adjoint representation of the gauge group, which we take to be 
$U(N)$.
We consider this theory in the presence or absence of superpotential interactions.
This is an interesting theory for the following reasons:
\begin{itemize}
\item[(1)] In the absence of superpotential deformations the theory is believed to be exactly 
superconformal for all $N,k$ \cite{Gaiotto:2007qi} and exhibits an R-symmetry with large quantum 
corrections. At strong coupling, $e.g.$ large 't Hooft coupling $\lambda=N/k\gg 1$ in the 
large-$N$ limit, the R-charge of the adjoint superfield tends to zero \cite{Niarchos:2009aa}. 
As a result, with increasing coupling more and more fields hit the unitarity bound and decouple
from the rest of the theory as free fields, thus truncating the interacting part of the chiral ring from below.
\item[(2)] In the presence of superpotential deformations and sufficiently strong coupling 
the theory exhibits spontaneous supersymmetry breaking \cite{Niarchos:2008jb}. 
In the range of parameters where a supersymmetric vacuum exists the superpotential deformation 
gives rise in the infrared (IR) to a new class of interacting fixed points. These theories are believed
to exhibit a Seiberg-like duality that acts on the theory in a self-dual manner \cite{Niarchos:2008jb}. 
Having a minimal matter content these examples provide some of the simplest illustrations of Seiberg 
duality in three dimensions and thus a useful playground for attempts to understand the general 
underpinnings of such dualities in field theory.
\end{itemize}

Our main goal will be to analyze the implications of these features on the properties of the free energy 
$F$. A preliminary analysis of $F$-maximization in these theories in the large-$N$ 't Hooft limit was 
presented in \cite{Niarchos:2011sn,Minwalla:2011ma}. Evaluating $F$ in the saddle point 
approximation it was confirmed numerically that $F$-maximization provides at $\lambda\ll N$ 
(but not necessarily $\lambda \ll 1$) results consistent with a set of non-perturbative inequalities
derived in \cite{Niarchos:2009aa}. This is a non-trivial test of the validity of the $F$-maximization 
principle. In section \ref{partition} we supplement this result with a more thorough examination of 
the solutions of the saddle point equations. Besides the one-cut solutions of Refs.\ 
\cite{Niarchos:2011sn,Minwalla:2011ma} we demonstrate the existence of a large class of multi-cut 
solutions and find numerical evidence that the one-cut solutions of Refs.\ 
\cite{Niarchos:2011sn,Minwalla:2011ma} are in fact the dominant ones (namely, they have the 
smallest free energy in the large-$N$ limit). The multi-cut solutions are expected to play an 
interesting role in the analysis of the non-perturbative (in $1/N$) instanton contributions to the 
matrix model that expresses the free energy after localization.

At $\lambda\sim N$ Ref.\ \cite{Niarchos:2011sn} observed numerically that one of the non-perturbative 
bounds of \cite{Niarchos:2009aa} gets violated by a naive application of the $F$-maximization 
principle. A potential source for this violation is the sizable number of free decoupled 
fields in this regime. In section \ref{modify} we test this expectation by suitably modifying the 
$F$-maximization principle to account for the decoupling fields. We find that the 
modification leads to an R-symmetry consistent with the non-perturbative bounds of 
\cite{Niarchos:2009aa}. We propose that the implemented modification is 
the general way to deal with the accidental symmetries of decoupling fields in $F$-maximization.

Another missing link in the $F$-maximization prescription is the following. Ref.\
\cite{Jafferis:2010un} argues for $F$-extremization, but the more precise statement about maximization
(namely the sign of the second derivative of $F$ with respect to the trial R-charges) is currently mostly 
an empirical observation rather than a generally argued fact.\footnote{Besides the explicit verification of 
maximization in specific examples, a general relation between $F$-maximization and 
volume-minimization has been proposed for large-$N$ CSM gauge 
theories with candidate Sasaki-Einstein gravity duals \cite{Martelli:2011qj,Cheon:2011vi,
Jafferis:2011zi}. See also the more recent work \cite{Gabella:2011sg}.} In the one-adjoint theories
that we consider we always observe numerically a single maximum that reproduces correctly in 
appropriate regimes independent information about the exact R-symmetry.

In addition, in section \ref{partition} we observe regions in $(\lambda, R)$-space ($\lambda$ being
the 't Hooft coupling and $R$ the trial R-charge of the adjoint chiral superfield), where the one-cut 
solutions cease to exist. This effect occurs precisely when the supersymmetric vacuum of the physical 
(superpotential-deformed) theory disappears. In the special case of the Chern-Simons matrix model, 
that expresses the free energy of the topological $\NN=2$ Chern-Simons theory, we demonstrate that 
this effect is directly related to the vanishing of the full non-perturbative partition function.

This observation motivates the possibility of a {\it general} relation between zeros of 
the $Q$-deformed sphere partition function, which leads to localization, and 
spontaneous supersymmetry breaking. We formulate and further motivate this potential 
relation in very general terms in section \ref{susy} and appendix \ref{bargument}. 
As a full-fledged theorem this relation would provide a new non-perturbative constraint on 
spontaneous supersymmetry breaking in three-dimensional QFTs that supplements the known constraints arising from the Witten index.

In section \ref{dual} we consider the second theme outlined in the beginning, namely Seiberg duality.
In a simple non-topological case, that involves a $U(1)$ and a $U(2)$ CSM theory,
we are able to compute the free energies analytically and demonstrate that they match exactly in the 
dual theories. In more general cases, we find convincing numerical evidence for this 
matching in the large-$N$ limit using our saddle point solutions.\footnote{Preliminary signs of 
potential disagreement with Seiberg duality reported in \cite{Niarchos:2011sn} are not verified.} 
In subsection \ref{dualdiscuss} further aspects of the duality of interest are discussed in some length.

Finally, the one-adjoint CSM theories possess a large class of RG flows induced by superpotential 
interactions. In section \ref{Ftheorem} we outline this web of RG flows and the general predictions 
implied by the conjectured F-theorem. In subsection \ref{Ftheoremtests} we find numerical evidence for 
the validity of these predictions in a set of examples with increasing complexity, thus providing 
new tests of the conjectured F-theorem.

In two extra appendices (app.\ \ref{multisaddle} and \ref{defmatching}) we summarize for the benefit of 
the reader a set of useful results that are alluded to, but not explicitly analyzed, in the main text.

\vspace{-0.3cm}
\subsection*{Summary of notation} 
The one-adjoint CSM theories that we consider are characterized by two integers, $N$ the rank
of the $U(N)$ gauge group, and $k$ the level of the CS interaction.\footnote{We assume $k>0$. The
case of negative $k$ can be obtained by a simple parity transformation.} Following 
\cite{Niarchos:2009aa} 
we frequently denote the theories without any superpotential interactions as $\hat {\bf A}$. 
New interacting fixed points arise by deforming the action with the superpotential interactions 
$W_{n+1}=\tr X^{n+1}$, $n=1,2,\ldots$.
We will variably denote the resulting theories either as ${\bf A}_{n+1}$ or in more detail as
$U(N)_k^{(n+1)}$.

A large-$N$ 't Hooft-like limit is possible and defined as 
\beq
\label{introab}
N, ~k \to \infty~, ~~ \lambda=\frac{N}{k}={\rm fixed}
~.
\eeq
In Refs.\ \cite{Niarchos:2009aa,Niarchos:2011sn} it was argued that as we increase the coupling 
$\lambda$ in the undeformed theory $\hat {\bf A}$ the R-charge function $R(\lambda)$ decreases 
monotonically towards zero attaining the values 
\beq
\label{introac}
R(\lambda^*_{n+1})=\frac{2}{n+1}~, ~~ n=1,2,\ldots
\eeq
at the critical couplings $\lambda^*_{n+1}$. At each of these points the operator $\tr X^{n+1}$ is, by 
definition, marginal. Using $F$-maximization, \cite{Niarchos:2011sn,Minwalla:2011ma}, one finds for 
the first few values of $n$ the values of $\lambda_{n+1}^*$ listed in Table \ref{list} above.

\begin{table}[t!]
\centering
\begin{tabular}{|c||c|c|c|c|c|c|c|c|c|c|} \hline
 $n$ & 3 & 4 & 5 & 6 & 7 & 8 & 9 & 10 \\ 
 \hline
 $\lambda_{n+1}^*$ & 0 & 0.35 & 0.65 & 0.9 & 1.15 & 1.4 & 1.7 & 2  \\
 \hline
\end{tabular}
\bf\caption{\it \small The numerically determined values of $\lambda^*_{n+1}$ using $F$-maximization
{\rm \cite{Niarchos:2011sn}} for $n=3,4,\ldots,10$.}
\label{list}
\end{table}

It has been conjectured that the ${\bf A}_{n+1}$ theories exhibit a Seiberg-like duality 
\cite{Niarchos:2008jb}. In the RG flows that give rise to the ${\bf A}_{n+1}$ theories the deforming 
operator $\tr X^{n+1}$ is relevant in both the `electric' and `magnetic' versions of the theory when 
$\lambda\in [\lambda^*_{n+1},n-\lambda^*_{n+1}]$. 
Following the nomenclature of analogous situations in four-dimensional gauge theories, we will 
sometimes call this interval the {\it `conformal window'} of the ${\bf A}_{n+1}$ theory. A review of the 
precise details of Seiberg duality in the ${\bf A}_{n+1}$ theories appears in section \ref{dual}.

%%%%%%%%%%%%%%%%%%%%%%%%%%%%%%%%%%%%%%%%
\section{$S^3$ partition function in the large-$N$ limit}
\label{partition}

The main object of interest in this paper is the $S^3$ partition function $Z_{S^3}$. Using localization
techniques \cite{Kapustin:2009kz,Jafferis:2010un,Hama:2010av} one can re-express $Z_{S^3}$ as 
a matrix integral. In our case, this matrix integral reads
\beq
\label{partitionaa}
Z_{S^3}(N,k,R)=\frac{1}{N!} \int \left( \prod_{j=1}^N e^{{\tt i}\pi k t_j^2} dt_j \right) 
\prod_{i<j}^N (2 \sinh(\pi t_{ij}))^2 \prod_{i,j=1}^N e^{\ell(1-R+{\tt i} t_{ij})}
=e^{-\FF(N,k,R)}
\eeq 
where $t_i$ are matrix eigenvalues running over the real line and $t_{ij}\equiv t_i-t_j$.
$\ell(z)$ is the function
\beq
\label{partitionaaa}
\ell(z)=-z\log\left( 1- e^{2\pi {\tt i} z} \right)+\frac{{\tt i}}{2} \left[ \pi z^2+\frac{1}{\pi} 
{\rm Li}_2 \left( e^{2\pi {\tt i} z} \right) \right] -\frac{{\tt i}\pi}{12}
~.
\eeq 
The free energy that we maximize with respect to $R$ is
\beq
\label{partitionab}
F=\frac{1}{2}(\FF+\bar \FF)
~.
\eeq

In the ${\bf A}_{n+1}$ theories the R-charge is fixed by the superpotential interaction 
$W_{n+1}=\tr X^{n+1}$, so in order to compute the free energy $F$ in these theories 
one simply sets $R=\frac{2}{n+1}$ in the expression \eqref{partitionaa}.

\subsection{Saddle point equations and their solutions revisited}
\label{saddle}

In the large-$N$ limit the main contribution to $Z_{S^3}$ comes from saddle point configurations
that obey the system of algebraic equations
\beq
\label{saddleaa}
\II_i \equiv \frac{\tt i}{\lambda}t_i +\frac{1}{N} \sum_{j\neq i}^N 
\left[ \coth (\pi t_{ij})-\frac{(1-R)\sinh(2\pi t_{ij})+t_{ij} \sin(2\pi R)}{\cosh(2\pi t_{ij})-\cos(2\pi R)} \right]
=0~, ~~ i=1,2,\ldots, N
~.
\eeq
At a saddle point configuration
\beq
\label{saddleab}
-\FF(\lambda, N)=-\log N!+
\sum_{i=1}^N \frac{{\tt i}\pi N}{\lambda} t_i^2 + \sum_{i<j}^N \log \left( 4\sinh^2 (\pi t_{ij}) \right)
+\sum_{i,j=1}^N \ell(1-R+{\tt i} t_{ij})
~.
\eeq
In general, the $N$ $t_i$'s that solve these equations are complex numbers (here $\C$-valued functions
of the parameters $R,\lambda$).

In lack of an efficient analytic method of solving these equations Refs.\ 
\cite{Niarchos:2011sn,Minwalla:2011ma} proceeded to analyze 
them numerically. In Ref.\ \cite{Niarchos:2011sn} solutions were found by introducing a fictitious time 
coordinate $\tau$ (following \cite{Herzog:2010hf}) and considering the dynamical evolution described 
by the set of differential equations
\beq
\label{saddleac}
a \frac{dt_i}{d\tau}=\II_i
~.
\eeq
In these equations $a=e^{{\tt i} \pi \theta}$ ($\theta\in [0,2\pi)$) is a free constant whose choice 
affects (in some cases) the solution that one converges to. Another crucial input that affects the 
obtained solution are the initial conditions at $\tau=0$. Using two independently written codes in 
Mathematica and Fortran we searched for different solutions varying $a$ and the initial conditions. 

\begin{figure}[t!]
\centering
\includegraphics[height=4.6cm]{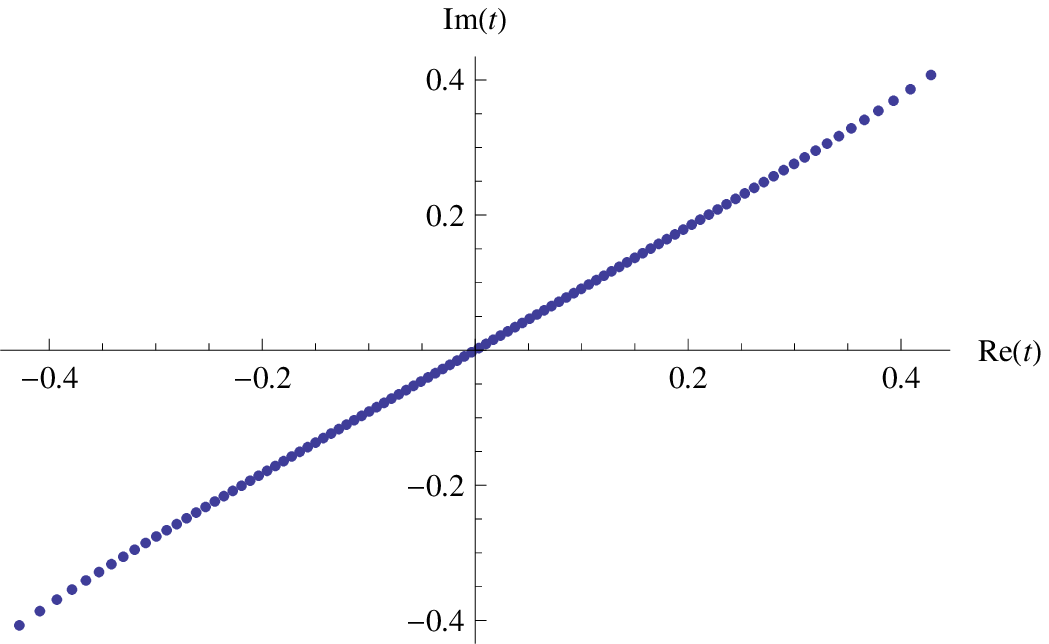} $(a)$ \hspace{0.1cm}
\includegraphics[height=4.6cm]{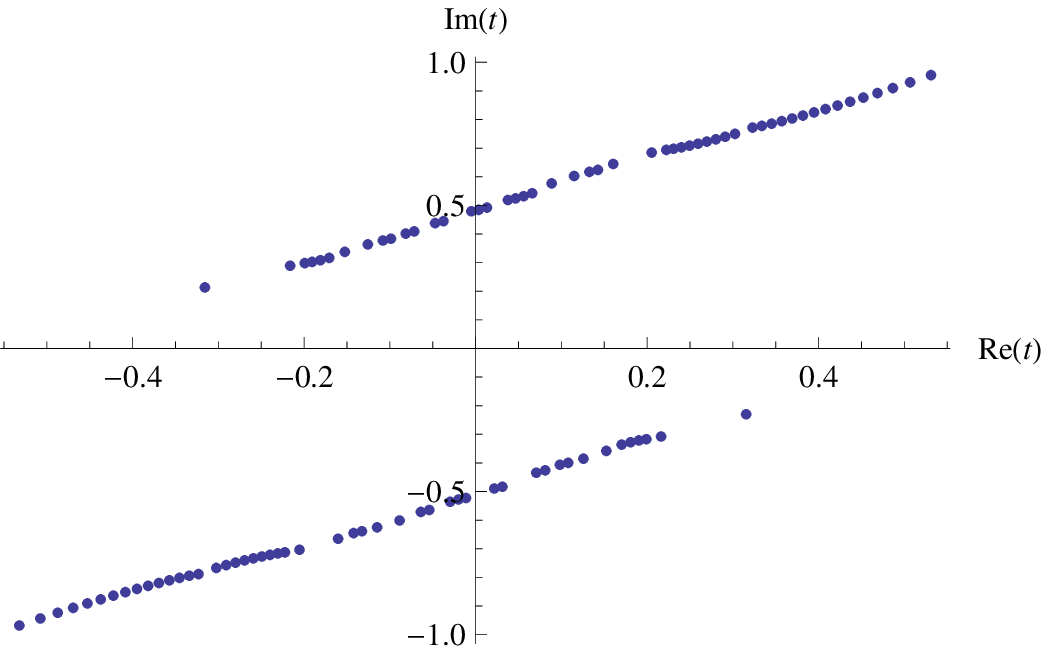} $(b)$
\bf\caption{\it \small Plot $(a)$: A typical one-cut distribution of the eigenvalues 
$t_i$ $(i=1,\ldots, N)$ in the complex plane. This particular plot has been obtained for $N=100$, 
$\lambda=1$ and $R=0.28$. Plot $(b)$: A typical two-cut solution depicted here for $N=100$, 
$\lambda=1.5$ and $R=0.5$. The specific example crosses the imaginary axis at $\pm \frac{\tt i}{2}$.}
\label{eigen}
\end{figure}

Besides the one-cut solutions that were reported in Refs.\ \cite{Niarchos:2011sn,Minwalla:2011ma} 
we have been able to detect many more multi-cut solutions with branches crossing the imaginary axis 
through the points $\pm \frac{m {\tt i}}{2}$, $\pm \frac{(R+m){\tt i}}{2}$, $(m=1,2,\ldots)$. A typical 
one-cut solution appears in Fig.\ \ref{eigen}$(a)$ and a typical two-cut solution in Fig.\ \ref{eigen}$(b)$. 
The multi-cut solutions can also be traced analytically in the weak coupling regime ($\lambda \ll 1$) 
with a perturbative computation (some of the key details of this computation are summarized in 
appendix \ref{multisaddle}).

The one-cut solutions were the ones used for $F$-maximization in Refs.\ 
\cite{Niarchos:2011sn,Minwalla:2011ma}, hence it is important to know if they are also the dominant 
ones in the large-$N$ limit ($i.e.$ if they are the ones having the minimal free energy for fixed 
$\lambda, R$). We confirmed this fact numerically in all cases that we checked. Further credence 
in the dominance of the one-cut solutions is provided by the following facts: $(i)$ at weak coupling 
they are the ones that reproduce the perturbative field theory computation of the R-charge 
\cite{Minwalla:2011ma}, $(ii)$ at finite coupling they reproduce to a good accuracy the predictions 
of Seiberg duality and F-theorem (see below).

The multi-cut solutions are expected to play an interesting role beyond the saddle point approximation.
In general, multi-cut solutions are closely related to the non-perturbative instanton corrections
that control the large-order behavior of the $1/N$ expansion \cite{Marino:2007te,
Marino:2008ya,Marino:2008vx,Pasquetti:2009jg}. Since such effects are beyond the 
immediate scope of the present work in what follows we will concentrate mostly on results based 
on the one-cut solutions.

%%%%%%%%%%%%%%%%%%%%%%%%%%%%%%%%%%%%%%%%%%%%%%%%%
\subsection{One-cut solutions and the functional behavior of $F(\lambda,R=\frac{2}{n+1})$ in the 
${\bf A}_{n+1}$ theories}
\label{1cutbreak}

Now we come to a more interesting observation. We focus on the ${\bf A}_{n+1}$ theories and
fix the R-charge to the value dictated by the deforming superpotential, hence $R=\frac{2}{n+1}$.
We would like to understand the functional behavior of the localized free energy function 
$F(\lambda,\frac{2}{n+1})$ as we vary the 't Hooft coupling $\lambda$. In particular, we would
like to understand what happens to this function when $\lambda>n$, $i.e.$ when the supersymmetric
vacuum is lifted. In that case the $Q$-deformed $S^3$ partition function computed via localization 
is no longer equal to the physical partition function of the ${\bf A}_{n+1}$ theory, but we can consider 
it anyway 
(for example, it gives information about the partition function $Z_{S^3}(\lambda, R)$ of the undeformed 
$\hat {\bf A}$ theory before $F$-maximization at $R=\frac{2}{n+1}$ ---see below for more comments 
related to this aspect).

A particularly instructive example that is amenable to full analytic control is provided by the ${\bf A}_2$
theory, $i.e.$ the case with $n=1$. In that case, the chiral superfield $X$ is massive and the theory flows 
in the IR to the topological $\NN=2$ Chern-Simons theory. The sphere partition function 
\eqref{partitionaa} simplifies considerably and can be computed exactly for any $N$, $k$.
Setting $R=1$ in \eqref{partitionaa} we obtain (using the identity $\ell(-z)=-\ell(z)$, $z\in \C$)
\beq
\label{saddleag}
\left | Z_{S^3}(\lambda, R=1)\right |=\frac{1}{N!} \left |  \int \left( \prod_{j=1}^N e^{{\tt i}\pi k t_j^2} dt_j \right) 
\prod_{i<j}^N (2 \sinh(\pi t_{ij}))^2 \right |
= \frac{1}{k^{N/2}} \prod_{m=1}^{N-1} \left ( 2\sin \frac{\pi m}{k} \right)^{N-m}
~.
\eeq
For the second equality we used a result that is explained, for example, in appendix B of Ref.\ 
\cite{Kapustin:2009kz}. Eq.\ \eqref{saddleag} gives the absolute value of the partition function of a
well-studied matrix model (the $U(N)$ Chern-Simons matrix model \cite{Marino:2002fk})
\beq
\label{saddleaga}
Z_{\rm CS}=\frac{1}{N!} \int \left( \prod_{j=1}^N e^{-\frac{2\pi^2 N}{S}t_j^2} dt_j \right)
\prod_{i<j}^N \left( 2 \sinh(\pi t_{ij})\right)^2 
~. 
\eeq
In the conventional case $S$ is a real parameter. In our case $S=2\pi {\tt i}\lambda$.

We would like to draw attention to the following property of the expression \eqref{saddleag}: 
the exact function $Z_{S^3} (\lambda,R=1)$ vanishes identically in the supersymmetry-breaking 
regime, $N>k$.\footnote{This fact was also observed in Ref.\ \cite{Kapustin:2010mh}.}
Accordingly, in this regime the free energy $F$, derived by localization, diverges. 
Since our information about the general $n$ case is currently limited to the leading-order large-$N$ 
limit, it is worth elaborating further on how this divergence exhibits itself in the saddle point 
approximation.

One way to derive the free energy $F$ at leading order in $1/N$ in the 't Hooft limit is by 
directly taking the large-$N$ limit of the exact expression \eqref{saddleag}. By converting a 
sum to an integral we obtain  
\beq
\label{saddleai}
f=-\frac{1}{N^2}\log |Z_{S^3}| = \int_0^1 d\mu (\mu-1) \log(2 \sin (\pi \lambda \mu))
~.
\eeq
For $\lambda\leq 1$ this integral evaluates to 
\beq
\label{saddleaia}
f=-\frac{\pi {\tt i}}{12\lambda} (1-3\lambda+2\lambda^2)-\frac{1}{4\pi^2 \lambda^2}
\left( {\rm Li}_3\left(e^{-2\pi {\tt i}\lambda} \right)-\zeta (3) \right)
~.
\eeq
Although not immediately obvious, the rhs of this equation is a real number.
For $\lambda>1$, $i.e.$ in the SUSY-breaking regime, part of the integrand in \eqref{saddleai} lies
on a logarithmic branch cut and the expression becomes ill-defined. 

Notice that $f$ vanishes at the SUSY boundary value $\lambda=1$. Hence, the crossover from
a well-defined regular $f$ at $\lambda\leq 1$ to a diverging $f$ at $\lambda>1$ happens 
discontinuously (more comments on the nature of this discontinuity below). The vanishing of 
$f$ at the general SUSY boundary values $N=k$ is a property of the exact $|Z_{S^3}|$ for any 
$N$, as can be seen from eq.\ \eqref{saddleag}. This generic feature in the ${\bf A}_{n+1}$ theory is 
a natural property given the vanishing of the rank of the gauge group in the Seiberg dual theory.

Another derivation of eq.\ \eqref{saddleai} can be obtained by solving directly  the saddle point  
equations. An analytic solution of these equations is known \cite{Aganagic:2002wv,Halmagyi:2003ze}.
For real $S$ in \eqref{saddleaga} the solution is a one-cut solution along the real axis with the 
eigenvalues occupying the interval
\beq
\label{saddleaib}
\CC=[-c,c]~, ~~ c=\frac{1}{\pi}{\rm arccosh}\left(e^{S/2} \right)
~.
\eeq
The resolvent of the solution is 
\beq
\label{saddleaic}
\omega(t)=2\log \left[ e^{-\frac{S}{2}} e^{\pi t} \left( \cosh(\pi t)- {\tt i} \sqrt{e^S-\cosh^2(\pi t)} \right)\right]
~,
\eeq
and the eigenvalue density
\beq
\label{saddleaid}
\rho(t)=-\frac{1}{\pi} {\rm Im}\, \omega(t)
=-\frac{1}{\pi {\tt i}}\log \frac{\cosh(\pi t)-{\tt i}\sqrt{e^S -\cosh^2 (\pi t)}}
{\cosh(\pi t)+{\tt i}\sqrt{e^S -\cosh^2 (\pi t)}}
~.
\eeq
By analytically continuing this result to $S\to 2\pi {\tt i}\lambda$ we obtain the one-cut solution 
that we are interested in. From the endpoint of the cut at 
\beq
\label{saddleade}
c=\frac{1}{\pi}{\rm arccosh}\left(e^{\pi {\tt i}\lambda} \right)
\eeq
it is apparent that as we increase $\lambda$ from 0 to 1, the cut rotates away from the real axis towards
the imaginary axis. At $\lambda=1$ the cut lies fully along the imaginary axis. Including the complex 
conjugate configurations that correspond to $\lambda\in [-1,0]$ one covers the full range of solutions
provided by the analytically continued version of \eqref{saddleaic}. In other words, there are no
one-cut solutions that can be identified in this way for $|\lambda|>1$, which corresponds to the 
non-supersymmetric regime of the ${\bf A}_2$ theory.

\begin{figure}[t!]
\centering
\includegraphics[height=4.6cm]{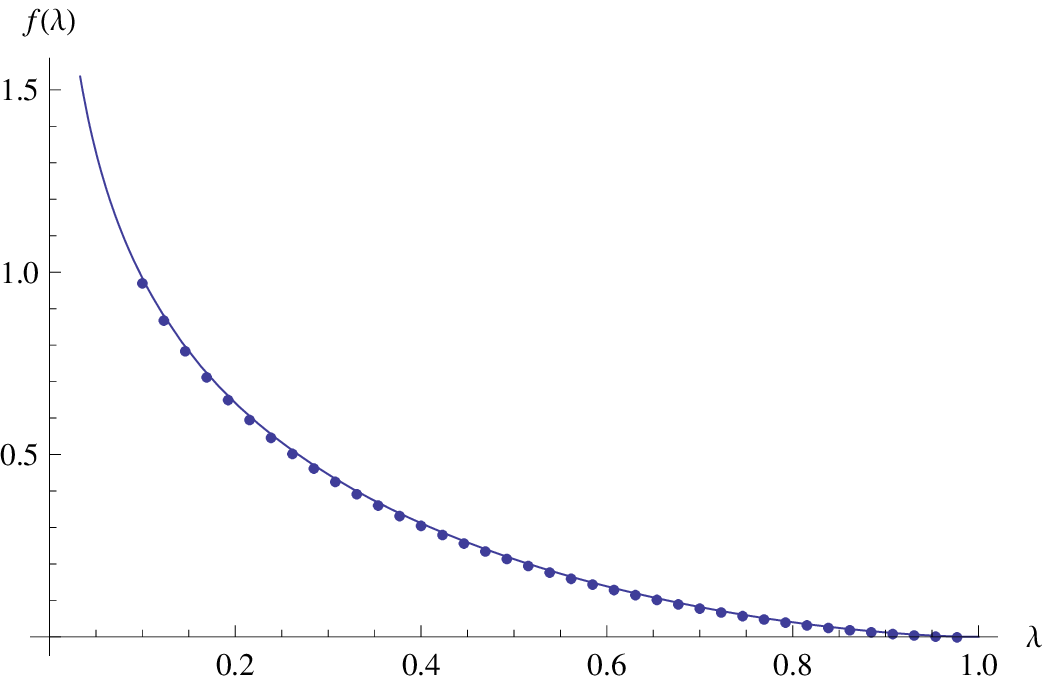} $(a)$ \vspace{0.2cm}
\includegraphics[height=4.6cm]{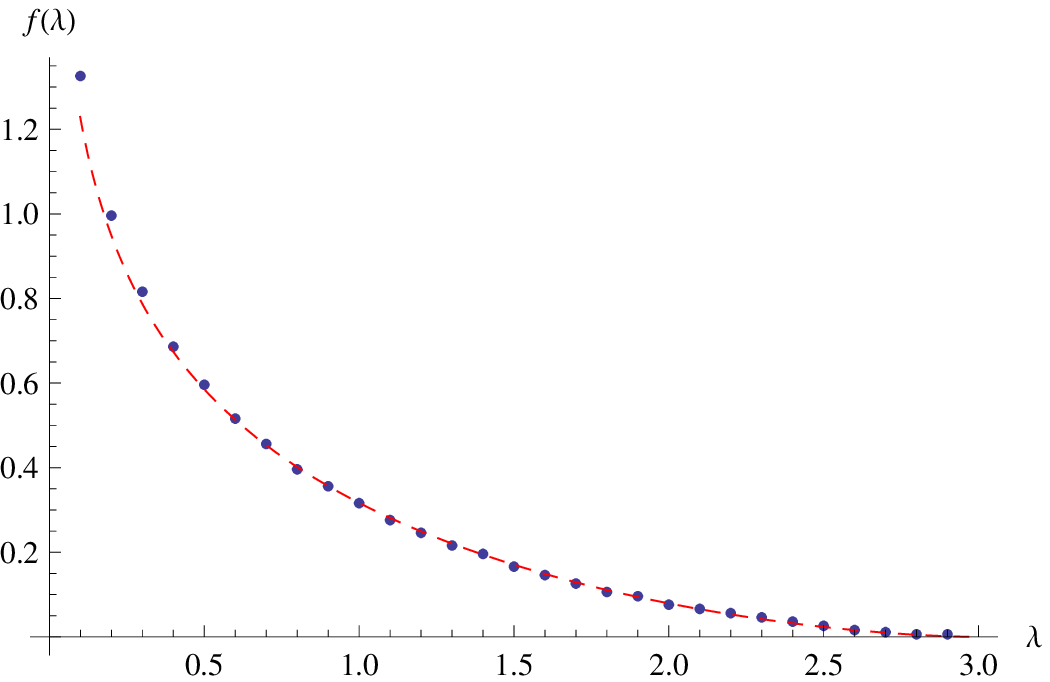} $(b)$
\bf\caption{\it \small Plot $(a)$ depicts the numerically determined renormalized free energy 
$f(\lambda)$ for the topological ${\bf A}_2$ theory (blue data points) in the supersymmetric interval
$[0,1]$. The blue curve depicts the analytic result \eqref{saddleai}. Plot $(b)$ provides the
analogous numerical result for the ${\bf A}_4$ theory (blue data points) for $\lambda\in[0,3]$. 
The dashed red curve is a fit provided by the expression \eqref{saddleaj}.}
\label{Flambda}
\end{figure}

This picture is fully reproduced by the numerical results of the previous subsection. We reproduce
the specifics of the eigenvalue distribution, $e.g.$ the $\lambda$-dependence of the one-cut endpoint
in \eqref{saddleade}, and the functional dependence of the free energy dictated by the 
expression \eqref{saddleai} for $\lambda \in [0,1]$ (see Fig.\ \ref{Flambda}$(a)$).\footnote{One
can view the good comparison of these data with the analytically expected result \eqref{saddleai}
as an additional test of the accuracy of our numerical results at $N=100$.} 
When we try to extend the numerical computation above $\lambda=1$ 
we fail to find any regular one-cut solutions and are instead able to obtain only multi-cut solutions. 
In view of the exact result \eqref{saddleag}, it is natural to associate the breakdown of the one-cut 
solutions at $\lambda>1$ to the non-perturbative divergence of $F$.

From a large-$N$ matrix model point of view one anticipates the following picture. In different regions
of moduli space the matrix model is dominated by different saddle points. In cases where a one-cut
phase dominates the corrections coming from multi-cut phases are exponentially small. In other 
cases, however, different multi-cut phases are of the same order, they sum in a coherent way and
the partition function can have zeroes. This type of asymptotics and phase transitions were studied in 
Ref.\ \cite{Marino:2009dp}. A more thorough discussion on the vanishing locus of the partition function 
can be found in the unpublished lecture notes \cite{marinoNotes}.\footnote{In \cite{marinoNotes} it is 
pointed out that zeroes of the partition function appear in many statistical systems as generalization of 
Lee-Yang zeroes. By the general theory of asymptotic expansions, they condense along anti-Stokes 
lines and in matrix models they are related to the zeroes of the $\vartheta$-function (see, for example, 
the footnote in page 39 of Ref.\ \cite{marinoNotes}). We are grateful to Marcos Mari\~no for explaining 
these aspects to us and for drawing our attention to the Refs.\ \cite{Marino:2009dp,marinoNotes}.}

Having a detailed understanding of the $n=1$ case, we can now proceed to ask what happens
at generic $n$ where our knowledge is unfortunately restricted only to a numerical solution of the 
large-$N$ saddle point equations. 

Our data indicate that the picture is qualitatively similar at all values of $n$. Nice one-cut
solutions exist throughout the whole supersymmetric interval, $\lambda \in [0,n]$, but disappear
discontinuously when the supersymmetric vacuum of the ${\bf A}_{n+1}$ theory is lifted. Extending the 
lessons of the $n=1$ case on the basis of the general picture outlined above,
we propose that this breakdown is always due to the non-perturbative 
divergence of $F$ in the supersymmetry-breaking region. 

A specific example of the numerically determined $\lambda$-dependence of the free energy $f$ 
in the supersymmetric interval is depicted in Fig.\ \ref{Flambda}$(b)$ for the ${\bf A}_4$ theory.
There is no known exact expression for the localized free energy in this non-topological case. 
Interestingly, we observe that the function
\beq
\label{saddleaj}
f_{fit,n=3}=0.8 \int_0^1 d\mu (\mu-1) \log\left( 2 \sin \left(\frac{\pi \lambda \mu}{3} \right) \right)
\eeq
provides a rather good fit to the numerical data. 
%According to this expression, in the non-supersymmetric 
%regime, $\lambda>3$, we encounter again a logarithmic branch cut that makes the solution ill-defined.
Qualitatively similar results can be obtained also for other values of $n$ with a fit of the form
\beq
\label{saddleaja}
f_{fit,n}=c_n \int_0^1 d\mu (\mu-1)\log\left( 2\sin \left( \frac{\pi \lambda \mu}{n} \right) \right)
~,
\eeq
for an appropriately chosen $n$-dependent number $c_n$. An appealing feature of this functional
form is that it obeys the correct Seiberg duality relation \eqref{matchab} that will be discussed later in 
section \ref{dual}. Unfortunately, despite this property, \eqref{saddleaja} cannot be the exact expression 
of the matrix model free energy at leading order in $1/N$ for generic $n$. At small $\lambda$ it can be 
checked that $f(\lambda, R)$ has the expansion detailed in eq.\ \eqref{saddleae} below, which would 
imply according to \eqref{saddleaja} $c_n=1$ for all $n$. This would be inconsistent with the presented
numerical results. 

It remains an interesting open problem to determine analytically the precise expression of $f$ at 
any $n$ and also to verify non-perturbatively in the matrix model that $Z_{S^3}=0$ when $N>nk$.
As a related point, in Ref.\ \cite{Kapustin:2010mh} it was shown analytically in a $U(N_c)$ 
Chern-Simons-Matter theory with $N_f$ pairs of (anti)-fundamental chiral multiplets that the matrix 
model $Z_{S^3}$ is always vanishing in the supersymmetry-breaking regime $N_c>|k|+N_f$.

\begin{figure}[t!]
\centering
\includegraphics[height=4.6cm]{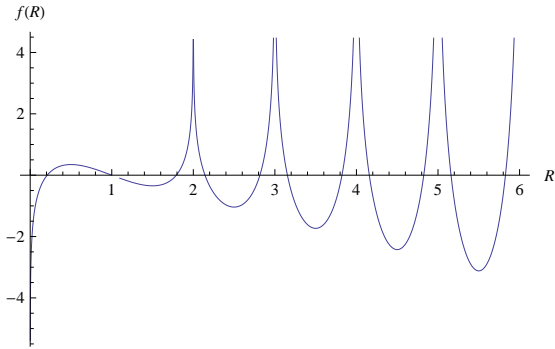}
\bf\caption{\it \small A plot of the free energy of a free field $f(R)=-\ell(1-R)$.}
\label{ell}
\end{figure}

We conclude this section with a couple of brief comments. 

The first comment has to do with the observed discontinuity of the free energy $F$ across
the SUSY-breaking boundary. Assuming $F$ (derived by localization) has to diverge in the 
SUSY-breaking region, the discontinuous jump from a finite value (in our set of examples $F=0$) 
to $+\infty$ is consistent with the intuition that the sphere free energy is a natural measure of the 
number of degrees of freedom. With increasing 
coupling gauge theories typically exhibit a reduction of the number of degrees of freedom, which is 
indeed in qualitative agreement with the observation that $F$ is a monotonically decreasing 
function of $\lambda$. In that sense, it would have been puzzling to observe that the free energy 
increases in the supersymmetric regime near $\lambda=n$ to meet the diverging behavior at 
$\lambda>n$ continuously.\footnote{We thank Andrei Parnachev for a question that prompted 
us to think about this aspect.}

The second comment is related to the general behavior of the function $F(\lambda, R)$. In
the context of $F$-extremization, it is interesting to know whether $F$ has a single or multiple 
maxima as a function of $R$ at fixed $\lambda$. Numerically we always find a single smooth maximum.
More specifically, at small fixed $\lambda\lesssim 0.2$ we observe that $F(R)$ exhibits a behavior similar to that of the free energy of a free field. Within perturbation theory one finds
\beq
\label{saddleae}
f(\lambda,R)\equiv \frac{1}{N^2} F(\lambda,R) = -\frac{1}{2}\log(4\pi^2 \lambda)-\ell(1-R)+\OO(\lambda^2)
~.
\eeq
The contribution $\ell(1-R)$ is the free energy of a free field depicted in Fig.\ \ref{ell}. 
As we increase $\lambda$ the behavior changes 
to one where the one-cut solutions cease to exist above a critical value of $R$.\footnote{We have
not been able to determine numerically the precise manner in which this change takes place as we 
vary $\lambda$. We consistently observed a critical value of $R$ for all the values of $\lambda>0.5$ that
we checked.} That critical value appears to be set by $R_{crit}(\lambda)=\frac{2}{\lambda+1}$. This is 
consistent with the previous discussion about the (non)existence of one-cut solutions in the 
${\bf A}_{n+1}$ theories.

%%%%%%%%%%%%%%%%%%%%%%%%%%%%%%%%%%%%%%%%%%%%%%%
\section{Spontaneous SUSY breaking and zeros of the localized $S^3$ partition function}
\label{susy}

The observations of the previous section suggest an intriguing relation between the $Q$-deformed
$S^3$ partition function that leads to localization and spontaneous supersymmetry breaking, 
similar in some respects to the one provided by the Witten index \cite{Witten:1982df}. Here we will 
attempt to formulate and motivate such a relation in a broad descriptive manner postponing a more 
rigorous treatment to future work.

Consider a general three-dimensional classically superconformal quantum field 
theory.\footnote{For reasons that will become obvious in a moment we assume that the theory has 
at least $\NN=2$ supersymmetry.}
Place the Euclidean version of this theory on a three-sphere $S^3$ and compute the partition function
\beq
\label{susyaa}
Z_{S^3}=\int e^{-S-t \int  \{ Q, V\} }
\eeq
where $S$ is the action of the theory, $Q$ is one of the supercharges and $V$ some  
fermionic functional. We assume $S$ is a properly defined action that leads to a correct
path integral treatment of the theory even at strong coupling.
$t$ is the coupling of the $Q$-exact deformation.
$V$ should be chosen in such a way that the deformation $\{ Q,V\}$
is positive definite and the standard localization technique can proceed without obstructions 
\cite{Pestun:2007rz}. For example, in the case of general Yang-Mills or Chern-Simons 
gauge theories with arbitrary matter content one can use the prescription of 
\cite{Kapustin:2009kz,Jafferis:2010un,Hama:2010av}.

In a theory with a supersymmetric vacuum a standard argument shows that the $t$-derivative 
of $Z_{S^3}(t)$ is identically zero. Hence, the physical $S^3$ partition function defined at
$t=0$ and its infinitely $Q$-deformed version at $t=\infty$ are identical
\beq
\label{susyab}
Z_{S^3}(0)=\lim_{t\to \infty} Z_{S^3}(t) \equiv Z_{S^3}^{(\rm loc)}
~.
\eeq
In cases where the standard localization techniques 
\cite{Kapustin:2009kz,Jafferis:2010un,Hama:2010av} (based on a UV Lagrangian formulation) can 
proceed unobstructed and capture correctly the quantum IR physics, the rhs of this equation, 
$Z_{S^3}^{(\rm loc)}$, can be further expressed in terms of a matrix integral formula. 
For general $\NN=2$ supersymmetric theories $Z_{S^3}^{(\rm loc)}$ is a 
function not only of the parameters of the theory ($e.g.$ the 't Hooft coupling $etc.$), but also of a set 
of trial R-charges. $F$-maximization postulates that the physical values of the R-charges can be 
determined by minimizing $|Z_{S^3}^{(\rm loc)}|$. So far, this is the familiar story that has led to the 
recent progress in three-dimensional supersymmetric quantum field theories.

$|Z_{S^3}^{\rm (loc)}|$, computed with this specific procedure, is the quantity one would naturally 
compute given a specific theory without any a priori knowledge about the fate of supersymmetry at 
the quantum level. In a theory with spontaneous supersymmetry breaking the $t$-derivative 
of $Z_{S^3}(t)$ is non-zero in general and the equality \eqref{susyab} is no longer valid
(see appendix \ref{bargument} for more details). For that reason the quantity 
$Z_{S^3}^{(\rm loc)}$ does not compute in this case the physical $S^3$ partition function of the 
theory $Z_{S^3}(0)$. Nevertheless, we would still like to claim that $Z_{S^3}^{(\rm loc)}$ is an 
interesting quantity to consider even when supersymmetry is spontaneously broken.

Our interest in this quantity stems from the potential validity of the following statements in a general 
classically superconformal quantum field theory (with at least $\NN=2$ supersymmetry):
\begin{itemize}
\item[$(a)$]{\it $Z_{S^3}^{\rm (loc)}=0$ implies that the theory exhibits spontaneous 
supersymmetry breaking.}
\item[$(b)$]{\it Spontaneous supersymmetry breaking implies $Z_{S^3}^{\rm (loc)}=0$.} 
\end{itemize}
Here we refer to spontaneous supersymmetry breaking on $S^3$. We believe this is 
ultimately related to spontaneous supersymmetry breaking on other three-manifolds, $e.g.$
$S^2\times \R$ which is directly related to the information carried by the Witten 
index \cite{Witten:1982df}. We will not, however, address this connection in this work.
Furthermore, since we refer to spontaneous supersymmetry breaking in a finite volume
we should mention that this does not necessarily imply that supersymmetry is also broken at infinite 
volume and the infinite volume limit may need to be treated with care \cite{Witten:1982df}. 

The first statement is natural for the following reasons. If the theory possesses a 
supersymmetric vacuum, $Z_{S^3}^{\rm (loc)}$ computes, as argued above, the physical partition
function on $S^3$. Then, assuming the free energy $F$ is a finite quantity for a 
quantum field theory with a finite number of degrees of freedom we conclude that 
$|Z_{S^3}^{\rm (loc)}|$ is a strictly positive quantity. The assumption that $F$ is finite is natural 
and expected to hold if $F$ defines a sensible analog of the $c$-function in three 
dimensions.\footnote{In particular, the application of the $F$-theorem would be subtle in RG flows
between theories with infinite $F$. In such cases $F_{UV}=F_{IR}=\infty$ and the validity of the
$F$-theorem, which states $F_{UV}>F_{IR}$, would be in general unclear.}
Therefore, the equation $Z_{S^3}^{\rm (loc)}=0$ cannot be true in a theory with a supersymmetric
vacuum and we should infer that, when it vanishes, $Z_{S^3}^{\rm (loc)}$ does not compute the 
physical partition function $Z_{S^3}(0)$. In that sense, the conclusion of spontaneous supersymmetry 
breaking seems to be a natural one. 

A potential pitfall should be noted here. Currently, the most efficient method to compute 
$Z_{S^3}^{\rm (loc)}$ proceeds via the localization technique which allows to recast the infinitely 
$Q$-deformed path integral \eqref{susyaa} in terms of a matrix integral. As we pointed out above, this 
method relies on the assumption that a certain UV Lagrangian description of the theory captures 
correctly the quantum IR physics. Incorrect conclusions may be reached if this extra assumption fails. 
For instance, one could encounter situations where a supersymmetric vacuum exists but the matrix 
integral computation gives $Z_{S^3}^{(\rm loc)}=0$. An example of this sort has been presented 
recently in independent work by Benini, Closset and Cremonesi \cite{Benini:2011mf}. The theory in 
question is three-dimensional super-QCD with gauge group $U(N_c)$ and $N_f=N_c-1$ pairs of 
(anti)fundamental multiplets. This theory has a quantum-mechanically deformed moduli space of 
supersymmetric vacua, but the naive computation of $Z_{S^3}^{(\rm loc)}$ via the standard matrix 
integral gives $Z_{S^3}^{(\rm loc)}=0$. Given the fact that the IR theory is a free theory of neutral 
chiral multiplets, one can argue that the actual value of $Z_{S^3}^{(\rm loc)}$ is non-zero and
equal to the value of the undeformed partition function in accordance with our general conjecture. 
The main lesson is that the sphere partition function $Z_{S^3}$ should be computed always, and at 
any deformation parameter $t$ including $t=0$ and $t=\infty$, using the proper formulation of the 
path integral.
 
The reverse statement $(b)$ could be motivated (or eventually argued) in a manner outlined in 
appendix \ref{bargument}. Under a set of assumptions outlined in appendix \ref{bargument} 
we argue that there are no states that can contribute non-trivially to the localized path integral
$Z_{S^3}^{\rm (loc)}$ when the theory exhibits spontaneous breaking of supersymmetry. 
This argument is qualitatively similar to the one showing that spontaneous 
supersymmetry breaking requires that the Witten index vanishes. As in the case of the 
Witten index, here also we would have been unable to reverse this argument to conclusively argue that
$Z_{S^3}^{\rm (loc)}=0$ implies spontaneous supersymmetry breaking (namely statement $(a)$).
It is interesting that, unlike the Witten index case, for $Z_{S^3}^{\rm (loc)}$ we can find an independent 
motivation for $(a)$ that goes through the finiteness of $F$ as a rough measure of degrees of freedom
and its relation to the $F$-theorem.

If correct, the combined statements $(a)$ and $(b)$ provide a new
powerful non-perturbative constraint on spontaneous supersymmetry breaking in three 
dimensions. Under the assumption that the standard localization techniques 
\cite{Kapustin:2009kz,Jafferis:2010un,Hama:2010av} proceed unobstructed,
they postulate that the free energy of a matrix model determines 
whether a corresponding 3D quantum field theory does or does not exhibit spontaneous 
supersymmetry breaking. The examples studied in this paper 
(and the examples in \cite{Kapustin:2010mh}) provide some evidence in favor of 
such constraints. The study of $Z_{S^3}^{\rm (loc)}$ in more examples with known patterns of 
spontaneous supersymmetry breaking may provide further insights into the fate of these statements. 
We hope to return to this task and a more formal argument in favor of $(a)$ and $(b)$ in a different 
publication.

Finally, we may add the following potentially interesting aspect to the above story.
In the context of statements $(a)$ and $(b)$ it may also be interesting to ask whether 
$Z_{S^3}^{\rm (loc)}$ combined with the 3D-4D connection, provides a way to relate supersymmetry 
breaking patterns in three and four dimensions. Similarities in such patterns are observed, 
for example, between the four-dimensional $\NN=1$ SQCD theory and 
three-dimensional $\NN=2$ CSM theories with fundamental and antifundamental matter, the 
four-dimensional $\NN=1$ one-adjoint SQCD theories and $\NN=2$ CSM theories of the type 
considered in this paper with additional matter in the (anti)fundamental representations $etc.$ 
\cite{Niarchos:2009aa}.

%%%%%%%%%%%%%%%%%%%%%%%%%%%%%%%%%%%%%%%%
\section{Decoupling fields and modifications of the $F$-maximization principle}
\label{modify}

It can be argued \cite{Niarchos:2009aa} that the exact $U(1)$ R-charge $R$ of the adjoint
chiral superfield $X$ in the $\hat {\bf A}$ theory tends to zero at strong coupling. For example, 
in the large 't Hooft coupling limit $\lim_{\lambda \to \infty} R(\lambda)=0$. In 
\cite{Niarchos:2011sn,Minwalla:2011ma} it was indeed shown 
with $F$-maximization that $R$ is a monotonically decreasing function of $\lambda$
that seems to asymptote to zero. 

As the R-charge decreases towards zero the scaling dimensions $\Delta_{n+1}=(n+1)R$ of the chiral 
ring operators $\tr X^{n+1}$ also decrease and, accordingly, with increasing coupling more and 
more chiral ring operators hit the unitarity bound and decouple as free fields. It is easy to show
\cite{Niarchos:2009aa} that the general operator $\tr X^{n+1}$ hits the unitarity bound for the first 
time when the operator $\tr X^{4(n+1)}$ becomes marginal. When a field decouples, a new accidental
symmetry occurs and the exact $U(1)$ R-symmetry of the theory can mix with it. In general, this
affects the validity and formulation of the $F$-maximization principle.

In the large-$N$, finite $\lambda \ll N$ computations of Refs.\ \cite{Niarchos:2011sn,Minwalla:2011ma} 
these effects were not important, because only a small finite number of free fields existed which had a 
negligible contribution to the total free energy of the theory that scales like $N^2$. However, as one 
increases $\lambda$ and makes it of the same order as $N$ the increased number of decoupled fields 
can have a sizable contribution to the free energy and the $F$-maximization principle must either be 
dropped or appropriately modified. The computation of these effects requires a direct analysis of the 
saddle point equations in the `M-theory' limit ---$N\to \infty$, $k$ finite--- which is unfortunately not a 
straightforward exercise. In the language of Ref.\ \cite{Jafferis:2011zi} this difficulty is due to the 
non-cancelation of long-range forces on the eigenvalues. On a more superficial level,
Ref.\ \cite{Niarchos:2011sn} observed numerically that the R-charge obtained by naive 
$F$-maximization at $\lambda\sim N=100$ violated a 
non-perturbative bound on the R-charge that follows from Seiberg duality and the assumption that $R$ 
is a monotonically decreasing function of $\lambda$. A  potential source of this discrepancy is
the considerable number of decoupling fields.

The general question of interest here is the following: is there a suitable modification of the $F$-
maximization principle that takes properly into account the effects of decoupling fields in a theory with 
large anomalous dimensions?  
In corresponding situations in four dimensions one is instructed to maximize a modified 
$a$-function where the 't Hooft anomalies associated to the free fields have been subtracted
\cite{Intriligator:2003jj,Kutasov:2003iy}. In this section we want to consider and test an analogous 
modification for $F$-maximization in three dimensions.

A natural course of action would simply be to subtract the free energy of the decoupling
fields. If $\OO$ is a decoupled operator with trial scaling dimension $\Delta$ its contribution to the 
free energy is
\beq
\label{modifyaa}
F_{free}=-\ell(1-\Delta)
~.
\eeq
 
Therefore, if $m$ chiral operators, $\tr X, \tr X^2,\ldots, \tr X^m$ have decoupled in our theory at 
't Hooft coupling $\lambda$, we propose that one should not maximize the free energy $F(\lambda,R)$ 
that follows from the matrix integral \eqref{partitionaa}, but the modified free energy
\beq
\label{modifyab}
F_{\rm mod}(\lambda, R;m)=F(\lambda,R)+\sum_{i=1}^m \ell(1-m R)
~.
\eeq

\begin{figure}[t!]
\centering
\includegraphics[height=5.2cm]{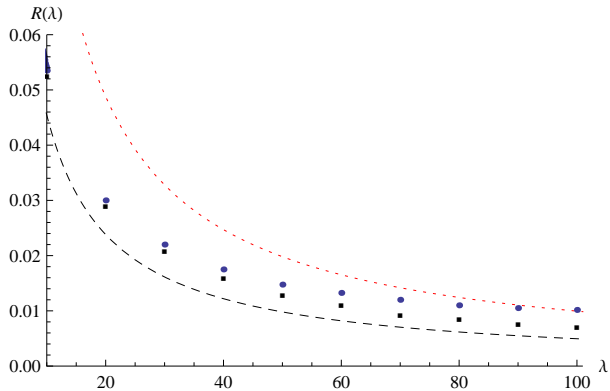}
\bf\caption{\it \small The numerically determined R-charge curve in the very strong coupling
regime for $10<\lambda<100$ and $N=100$. The upper blue points represent the result of 
the standard $F$-maximization recipe. The lower black square points represent the result of the 
modified $F$-maximization recipe \eqref{modifyab} which is designed to take into account the 
effects of the decoupling fields. The exact result is required by consistency to lie within the bounds 
of the lower dashed black curve $\frac{1}{2(\lambda+1)}$ and the upper dotted red curve 
$\frac{2}{2\lambda+1}$ (the lower bound holds for $\lambda \in \N$ and the upper bound for 
$\lambda \in \N/2$) {\rm \cite{Niarchos:2011sn}}.}
\label{strong}
\end{figure}

We implemented numerically this modified $F$-maximization principle to the $\hat {\bf A}$ theory and 
the result is plotted (for $N=100$) in Fig.\ \ref{strong}. At each $\lambda$ we determined $m$, the 
number of free fields, in a self-consistent manner by requiring $\tr X^m$ to be the highest in polynomial 
power chiral ring operator that is decoupled after $F_{\rm mod}$-maximization is implemented. 
In Fig.\ \ref{strong} the blue round dots are the result of the standard $F$-maximization procedure that 
exhibits the above-mentioned violation near $\lambda \sim 100$ ($i.e.$, the curve traced by the blue 
data points crosses the upper-bounding red curve given by $\frac{2}{2\lambda+1}$) 
\cite{Niarchos:2011sn}. The black square dots are 
the result of $F_{\rm mod}$-maximization. We observe that the modified principle brings down the value
of $R$ towards the black dashed lower-bound curve. The new R-charge curve lies between the upper
and lower bound curves and exhibits no signs of bound violation.\footnote{For a clear explanation 
of the origin and meaning of the bound curves we refer the reader to 
\cite{Niarchos:2009aa,Niarchos:2011sn}.} Moreover, a naive fit of the
black data points gives at large $\lambda$ the following asymptotic behavior of $R(\lambda)$
\beq
\label{modifyac}
R(\lambda) \sim 0.46\, \lambda^{-0.91}
\eeq
which agrees with the intuition that $R(\lambda)$ should trace closely (or asymptote to) the lower 
bounding curve given by $\frac{1}{2(1+\lambda)}$ \cite{Niarchos:2011sn}.

This short computation is suggestive of the validity of $F_{\rm mod}$-maximization, but unfortunately
does not provide conclusive evidence. The above numerical computation has been performed at
a finite, but large, value of $N$ where the leading order saddle point approximation is corrected by 
several sources of $1/N$ contributions. It is not completely clear how these $1/N$ effects compare with 
the contributions due to the decoupling fields. For that reason, it would be desirable to implement and 
test our proposal in more examples of CSM theories that exhibit decoupling fields, $e.g.$ examples 
that include extra fields in (anti)fundamental representations \cite{Niarchos:2009aa}. In such cases,
the contribution of decoupling meson-like operators would be of the same order as the 
leading order free energy.

%%%%%%%%%%%%%%%%%%%%%%%%%%%%%%%%
\section{Duality in ${\bf A}_{n+1}$ theories}
\label{dual}

In Ref.\ \cite{Niarchos:2008jb} evidence was provided for a Seiberg-like duality in ${\bf A}_{n+1}$ 
theories.\footnote{Ref.\ \cite{Niarchos:2009aa} considered generalizations of this duality in theories 
with one or two adjoint chiral superfields and additional chiral superfields in the (anti)fundamental 
representations. In Ref.\ \cite{Jafferis:2011ns} a duality was proposed for an $SU(2)$ $\NN=2$ CS 
theory at level one coupled to a single chiral superfield in the adjoint and no superpotential interactions.
A Seiberg duality in three dimensional Chern-Simons-Matter theories was originally proposed by
Giveon and Kutasov in \cite{Giveon:2008zn} for a theory with matter in (anti)fundamental 
representations.}
At finite $N$, $k$ the duality relates the $U(N)$ ${\bf A}_{n+1}$ theory at level $k$ to the $U(nk-N)$
theory at level $-k$; in convenient notation
\beq
\label{dualaa}
U(N)_k^{(n+1)} ~\leftrightarrow ~ U(nk-N)^{(n+1)}_{-k}
~.
\eeq
In this case Seiberg duality acts self-dually and exchanges only the rank of the gauge group and
the sign of the CS level. In the large-$N$ 't Hooft limit the duality acts by exchanging 
$\lambda \leftrightarrow n-\lambda$ (where here we define $\lambda=\frac{N}{|k|}$). 
In terms of the number of matter fields involved this is one of the simplest known dualities in three 
dimensions.

Let us briefly review the arguments in favor of this duality. Originally, this duality was motivated in 
Ref.\ \cite{Niarchos:2008jb} with a standard D-brane argument based on a type IIB string theory 
setup that involves $N$ D3-branes
suspended between $n$ NS5-branes and a single $(1,k)$ 5-brane bound state suitably oriented
to preserve the right amount of supersymmetry. In order to argue for duality one moves the fivebranes 
through each other and considers the effects of this motion on the D3-branes (which is the place
where the Chern-Simons-Matter theory of interest lives). To the degree that this brane motion 
is inconsequential for the IR physics, this is an argument in favor of duality in the IR field theory on
the D3-branes. A potentially subtle point in this procedure has to do with the fact that, 
because of the particular orientation of the branes, the fivebranes have to meet and 
cross each other in spacetime at a point where singular behavior
can arise triggering a transition. Such phenomena are known to occur, but the 
presence of the Chern-Simons interaction is believed to alleviate the singular behavior and help
the theory avoid a transition (see \cite{Kapustin:2010mh} for a relevant discussion and explicit 
examples).

In our case, additional evidence for the validity of the duality is provided by the following facts:
\begin{itemize}
\item[(1)] In the special case where $n=1$ the theory flows in the IR to the topological 
$\NN=2$ CS theory. Duality in this case reduces to level-rank duality and thus one can check 
explicitly the matching of the $S^3$ partition functions \cite{Kapustin:2010mh}.
\item[(2)] In the general $n$ case, we can further deform the superpotential to 
\beq
\label{dualab}
W=\sum_{i=0}^n \frac{g_{i}}{n+1-i} \tr X^{n+1-i}
\eeq 
and flow in the deep IR to a product of ${\bf A}_{n_i}$ theories $(n_i<n)$. 
Suitably arranging the coefficients $g_i$ one can flow to a product of ${\bf A}_2$ theories
\beq
\label{dualac}
U(N)_k^{(n+1)} \to U(N_1)_k^{(2)} \times \cdots \times U(N_n)_k^{(2)}~, ~~ \sum_{i=1}^n N_i=N
~.
\eeq 
As already noted, the duality holds in each of the $U(N_i)_k^{(2)}$ factors as level-rank duality.
One might expect that the duality continues to hold as we tune back all the coefficients
$g_i$ $(i=1,\ldots,n)$ to zero, keeping $g_0$ non-zero, in order to recover the original 
$U(N)_k^{(n+1)}$ theory. 
A more detailed discussion of these RG flows will appear in the 
next section \ref{Ftheorem}. The behavior of the duality under the general superpotential 
deformation is further discussed in appendix \ref{defmatching}.
\end{itemize}

In what follows we attempt to find more detailed tests of the duality using the localized 
$S^3$ partition function. The duality predicts that the $S^3$ partition function is invariant
under duality (up to an overall phase which is generally attributed to different 
framing \cite{Kapustin:2010mh,Witten:1988hf}). 
In our case, this statement translates to the following set of mathematical 
identities (for $N\leq nk$)
\bea
\label{dualad}
&&Z_{S^3}\left[ U(N)_k^{(n+1)} \right]=\frac{1}{N!} \int \left( \prod_{j=1}^N e^{{\tt i}\pi k t_j^2} dt_j \right)
\prod_{i<j}^N \left( 2\sinh(\pi t_{ij} )\right)^2 \prod_{i,j=1}^N e^{\ell \left (\frac{n-1}{n+1}+{\tt i}t_{ij} \right)}
=\nonumber\\
&&e^{{\tt i}\vartheta(N,k,n)} Z_{S^3}\left[ U(nk-N)_{-k}^{(n+1)} \right]
=\\
&&\frac{e^{i\vartheta(N,k,n)}}{(nk-N)!} \int \left( \prod_{j=1}^{nk-N} e^{-{\tt i}\pi k t_j^2} dt_j \right)
\prod_{i<j}^{nk-N} \left( 2\sinh(\pi t_{ij} )\right)^2 \prod_{i,j=1}^{nk-N} 
e^{\ell \left (\frac{n-1}{n+1}+{\tt i}t_{ij} \right)}
\nonumber
\eea
where $\vartheta(N,k,n)$ is some phase. In other words, the free energies $F$ are duality invariant. 
We will test the validity of these identities first exactly in a simple, but non-topological, low-$N$ case,
and then in the large-$N$ 't Hooft limit by using the results of our numerical computation. The 
matching in the latter case also provides supplemental evidence for the validity of these results
and the accuracy of the numerical computation. 

Another way to derive the duality invariance of the free energy is by reduction of a superconformal
index in four dimensions \cite{Dolan:2011rp}. Starting from the electric-magnetic duality of an
$Sp(2N)$ theory in \cite{Intriligator:1995ff} it is possible to reduce the matching of the corresponding
superconformal indices to the equality of three-dimensional partition functions in 
\eqref{dualad}.\footnote{We thank Grigory Vartanov for a discussion on this aspect.}

Other detailed aspects of the duality will be discussed in subsection \ref{dualdiscuss}.

\subsection{A simple non-topological duality}
\label{analytic}

One can readily check the validity of the identities \eqref{dualad} in the topological $n=1$ case.
The `hard' $e^{\ell(\ldots)}$ factors drop out in this case and the integrals can be computed 
straightforwardly by brute force (see $e.g.$ eq.\ \eqref{saddleag}). In general, one finds  
\cite{Kapustin:2010mh}
\beq
\label{analyticaa}
Z_{S^3}\left[ U(N)_k^{(2)} \right] = e^{-\frac{\pi {\tt i}}{12}(k^2-6k+2)}
Z_{S^3}\left[ U(k-N)_{-k}^{(2)} \right] 
~.
\eeq

Now let us consider a simple non-topological case. We set $N=2, k=1,n=3$ which leads to the duality
\beq
\label{analyticab}
U(2)_1^{(4)} \leftrightarrow U(1)_{-1}^{(4)}
~.
\eeq
The $U(1)$ partition function involves a single Gaussian integral that gives (by analytic continuation)
\beq
\label{analyticac}
Z_{S^3}\left[ U(1)_{-1}^{(4)} \right]=\frac{1}{\sqrt 2} e^{-\frac{\pi {\tt i}}{4}}
~.
\eeq
The $U(2)$ partition function involves a double integral
\beq
\label{analyticad}
Z_{S^3}\left[ U(2)_1^{(4)} \right]=\frac{1}{2} \int_{-\infty}^\infty dt_1 \, dt_2 \,
e^{{\tt i}\pi (t_1^2+t_2^2)} \frac{\sinh^2(\pi t_{12})}{\cosh(\pi t_{12})}
~.
\eeq
We used the fact that $e^{2\ell\left(\frac{1}{2} \right)}=\frac{1}{2}$. With a simple change of
integration variables we can factor out a Gaussian integral and reduce the computation to 
\beq
\label{analyticae}
Z_{S^3}\left[ U(2)_1^{(4)} \right]= \frac{1}{2\sqrt 2}e^{-\frac{3\pi {\tt i}}{4}}
\int_{-\infty}^\infty dt \, e^{\frac{\pi {\tt i}}{2} t^2} \frac{\sinh^2(\pi t)}{\cosh(\pi t)}
~.
\eeq
Using the identity $\sinh^2(\pi t)=\cosh^2(\pi t)-1$ we can further split this integral into a Gaussian 
integral and an integral that can be computed directly with Mathematica
\beq
\label{analyticaf}
Z_{S^3}\left[ U(2)_1^{(4)} \right]= \frac{1}{2\sqrt 2}e^{-\frac{3\pi {\tt i}}{4}}
\left( \sqrt{2} e^{\frac{3\pi {\tt i}}{4}}-
\left( 2 e^{-\frac{\pi {\tt i}}{8}}- \sqrt 2 e^{-\frac{\pi {\tt i}}{4}} \right) \right)
=\frac{1}{\sqrt 2} e^{\frac{\pi {\tt i}}{8}}
~.
\eeq
We conclude that 
\beq
\label{analyticag}
Z_{S^3}\left[ U(2)_1^{(4)} \right]=e^{\frac{3\pi {\tt i}}{8}}  Z_{S^3}\left[ U(1)_{-1}^{(4)} \right]
\eeq
which verifies the prediction made by the duality in this case.

Notice that the relative phase in \eqref{analyticag} is not the same as that in eq.\ \eqref{analyticaa}
for $k=1$. This implies that the relative phase has, in general, both a $k$ and an $n$ dependence.

The particular example we have just discussed demonstrates the important role of the $U(1)$ part 
of the gauge group in the duality. For an $SU(2)$ theory at level 1 and no superpotential interaction
a different type of duality was conjectured in \cite{Jafferis:2011ns}.

\subsection{Free energy matching tests at large-$N$}
\label{largeNmatch}

\begin{figure}[t!]
\centering
\includegraphics[height=4.5cm]{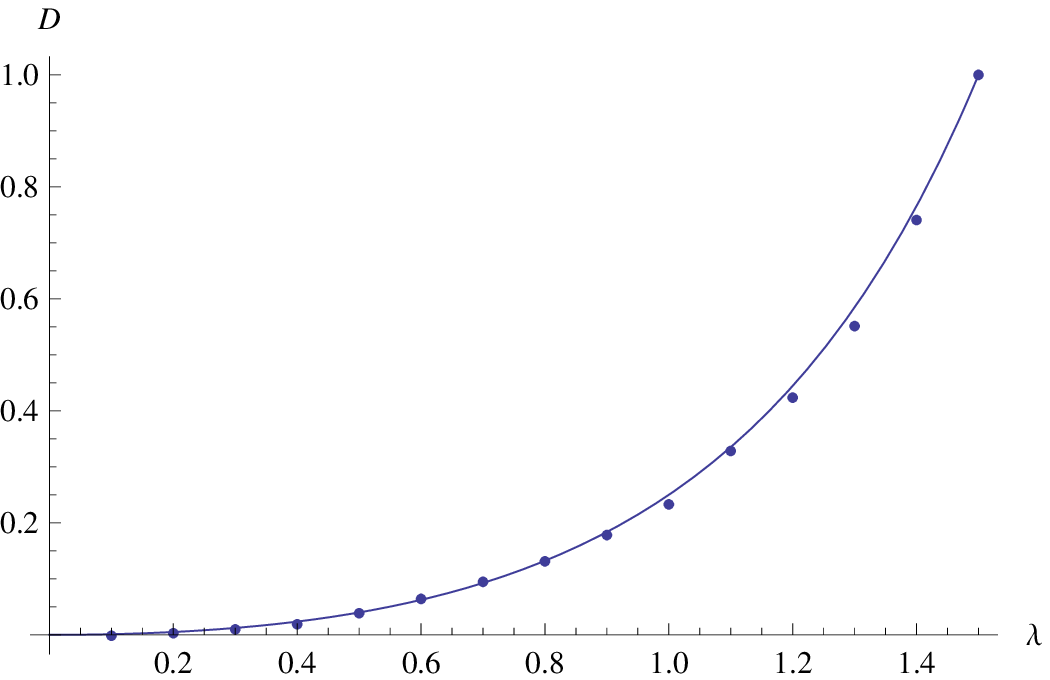} $(a)$
\includegraphics[height=4.5cm]{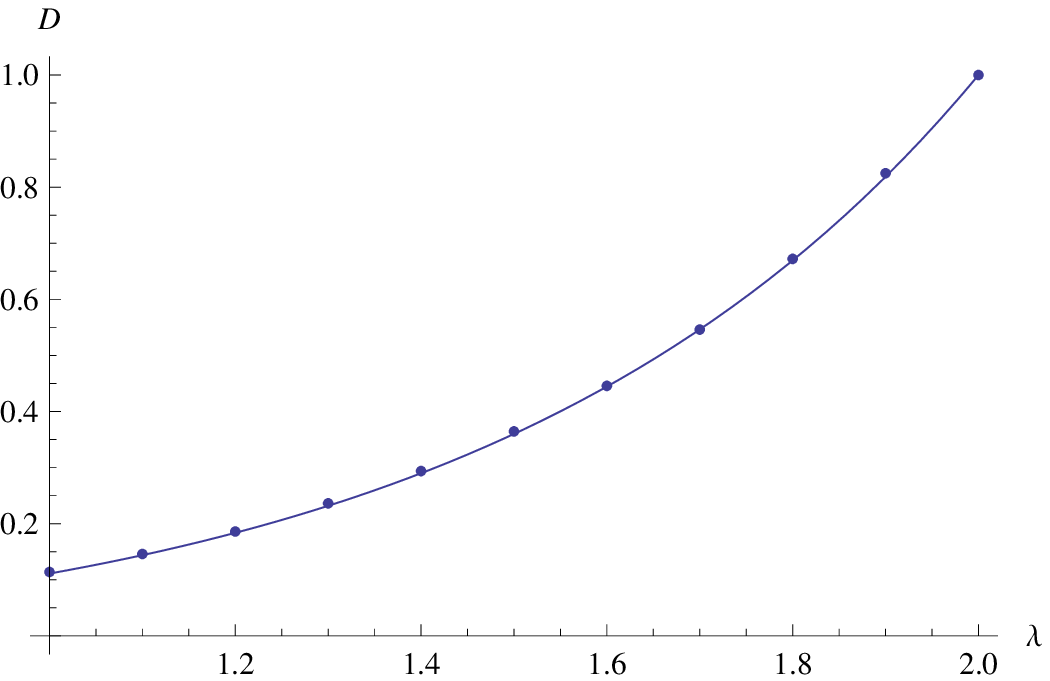} $(b)$
\includegraphics[height=4.5cm]{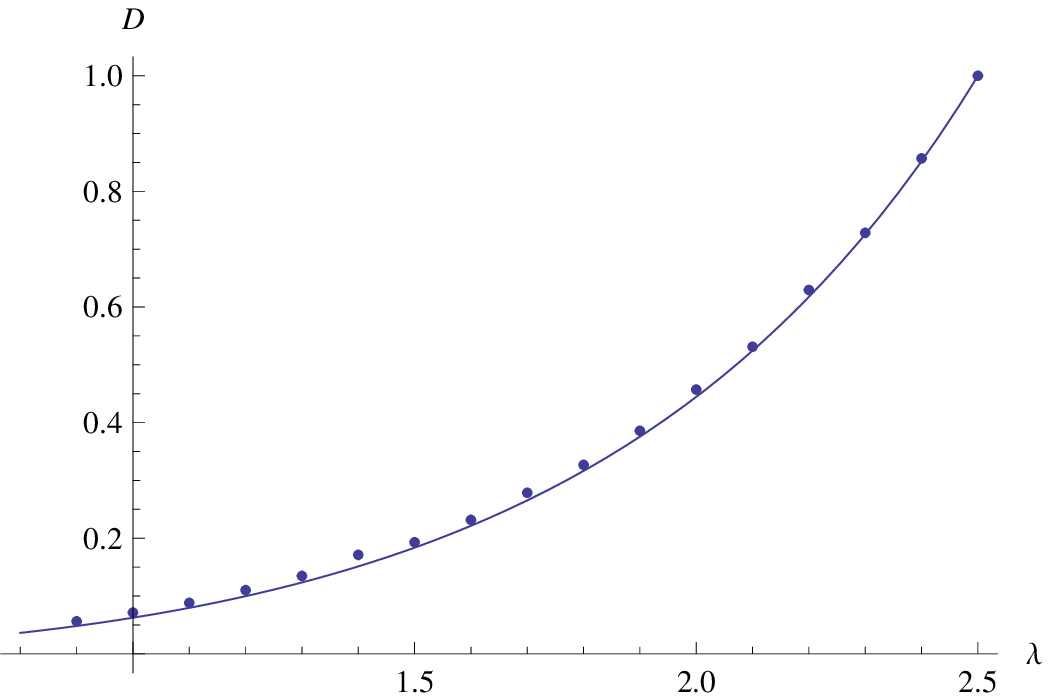} $(c)$
\bf\caption{\it \small Plots of the function $D(\lambda)$, defined in terms of the free energies
in eq.\ \eqref{matchac}, for the ${\bf A}_4$ theory (plot (a)), the ${\bf A}_5$ theory (plot (b)) and
the ${\bf A}_6$ theory (plot (c)). The blue points represent the numerically computed results and
the solid curves the duality-predicted functions $\lambda^2(n-\lambda)^{-2}$.}
\label{duality}
\end{figure}

We can provide additional tests of the above duality in the large-$N$ 't Hooft limit 
by using the numerical results of section \ref{partition}. Focusing on the planar part of
the free energy we define 
\beq
\label{matchaa}
f(\lambda)=\frac{1}{N^2} F(\lambda, N)
~.
\eeq
The duality predicts the equality
\beq
\label{matchab}
F(\lambda, N) = F\left (n-\lambda, \left( \frac{n}{\lambda}-1 \right) N \right) ~ \Leftrightarrow ~
f(\lambda)= \left( \frac{n}{\lambda}-1 \right)^2 f(n-\lambda)
~.
\eeq

In Fig.\ \ref{duality} we define
\beq
\label{matchac}
D(\lambda) \equiv \frac{f(n-\lambda)}{f(\lambda)}
\eeq
which according to the duality should be equal to $\lambda^2 (n-\lambda)^{-2}$. Using our
numerical data we present how the matching works in three cases: $n=3,4,5$ ($i.e.$ the 
${\bf A}_4,{\bf A}_5,{\bf A}_6$ theories). For $n=3$ the conformal window extends over the
whole supersymmetric interval $\lambda\in [0,3]$. For $n=4$ the conformal window extends
over the range $\lambda\in (0.35, 3.65)$. For $n=5$ the conformal window extends over the
range $\lambda \in (0.65, 4.35)$. In all cases we cover most of the conformal window fixing 
the probed range according to the ability of the numerical code to produce trustworthy results.
The plotted data terminate at the self-dual point of each theory which lies at $\lambda_{sd}=\frac{n}{2}$.
The agreement between the curve traced by the numerical
points and the curve $\lambda^2 (n-\lambda)^{-2}$ predicted by duality is rather convincing.

\subsection{Other aspects of duality}
\label{dualdiscuss}

The successful matching of the free energies $F$ in the above examples, combined with 
the arguments summarized in the beginning of this section, provide non-trivial evidence for 
the validity of the dualities \eqref{dualaa}. It would be very interesting to show analytically
the matching of the free energies \eqref{dualad} for general $N$, $k$ (modulo potential intricacies
that are discussed below). Transformation properties of hyperbolic hypergeometric integrals 
\cite{spiri1,spiri2,Rains,Bult} may be very useful in this respect (as they have been in other cases, 
$e.g.$ in the case of the Giveon-Kutasov duality \cite{Giveon:2008zn,Willett:2011gp}).
To the best of our knowledge the identities \eqref{dualad} have not been proven in this way
before (see \cite{Spiridonov:2009za} for the appearance of analogous identities in four-dimensional 
theories). In anticipation of such results we would like to offer at this point a few additional
comments on some particularly interesting, less discussed, physical aspects of the dualities 
\eqref{dualaa}. For simplicity, we will focus on the large-$N$ 't Hooft limit where the ${\bf A}_{n+1}$ 
theories are parameterized by $n$ and the continuous 't Hooft coupling $\lambda$.

We have argued (see \cite{Niarchos:2009aa} for a detailed discussion) that there is a critical coupling
$\lambda_{n+1}^*$ where the deforming operator $\tr X^{n+1}$ becomes marginal. For 
$\lambda<\lambda_{n+1}^*$ the deforming operator is irrelevant (for $n>3$) and the theory
erases the deformation in the deep IR where one recovers the undeformed $\hat {\bf A}$ theory. 
For $\lambda>\lambda_{n+1}^*$ the deforming operator is relevant and drives the theory to 
a new IR fixed point, which is the theory we are interested in. The global symmetry group of this
theory is a single $U(1)$, the R-symmetry group. The R-charge of the adjoint chiral superfield
is fixed by the superpotential 
\beq
\label{dualdiaa}
R(X)=\frac{2}{n+1}
~.
\eeq

The duality is expected to match symmetries and operators. Indeed, this works very well inside
the range of $\lambda$'s that we call the conformal window of this theory. This is the range where
the deforming operator $\tr X^{n+1}$ is relevant both in the `electric' $U(N)$ theory and the 
`magnetic' $U(nk-N)$ theory, namely when
\beq
\label{dualdiab}
\lambda \in [\lambda^*_{n+1},n-\lambda_{n+1}^*]
~.
\eeq
In this case, the $U(1)$ R-symmetry is identically the same on both sides of the duality because
it is controlled by the same superpotential interaction, $i.e.$
\beq
\label{dualdiac}
R(X)=R(\bar X)=\frac{2}{n+1}
\eeq
where $\bar X$ denotes the adjoint chiral superfield on the magnetic side. Notice, that because of
this, the number of decoupled free chiral operators is the same on both sides of the duality. 
Specifically, the single-trace operators
\beq
\label{dualdiad}
\tr X,~\tr X^2,~\cdots, \tr X^{\left[ \frac{n+1}{4} \right]}
\eeq
are free and decoupled on both sides of the duality ($[x]$ denotes the integer part of the number $x$).
Accordingly, under duality not only $F$ but also its modified version \eqref{modifyab} is invariant.

As an aside comment, we note the following interesting property of the RG flow from the UV undeformed
$\hat {\bf A}$ theory to the IR deformed theory ${\bf A}_{n+1}$ in this range of $\lambda$. In the 
UV the number of free decoupled fields is $m$ and this number is controlled by the exact R-symmetry
of the $\hat {\bf A}$ theory which is determined by $F$-maximization. Specifically,
\beq
\label{dualdiae}
m=\left[ \frac{1}{2 R(\lambda)} \right]
~.
\eeq
Along the RG flow from $\hat {\bf A}$ to ${\bf A}_{n+1}$ the R-charge increases and $m$ drops in 
general from \eqref{dualdiae} to $\left[ \frac{n+1}{4} \right]$, which implies that a number of UV-free
operators recouple in the IR and regain their interacting status. It would be useful to obtain a deeper 
understanding of the mechanism that realizes this effect.

Returning to the ${\bf A}_{n+1}$ theory one can do a bit better in matching the operators. 
Following \cite{Kutasov:1995ss} in appendix \ref{defmatching} we demonstrate how one 
matches the general 
superpotential deformation \eqref{dualab} and the associated vacuum structure on both sides of the 
duality. A simple corrollary of that analysis is the map
\beq
\label{dualdiaf}
\tr X^i \leftrightarrow - \tr \bar X^i ~, ~~ i=1,2,\ldots, n+1
\eeq
for electric superpotential $W=\tr X^{n+1}$ and magnetic superpotential $\overline{W}=\tr \bar X^{n+1}$.
Under this map, a free decoupled operator from the list \eqref{dualdiad} on the electric side maps
to a free decoupled operator from the corresponding list on the magnetic side.

This appears to be a satisfactory picture of how 3D Seiberg duality works in this class of 
Chern-Simons-Matter theories. Now one can ask what happens when $\lambda$ is increased
further and dialed to lie in the window
\beq
\label{dualdiag}
[n-\lambda^*_{n+1}, n]
\eeq
where a supersymmetric vacuum continues to exist, but the deforming operator is relevant in 
the electric theory and {\it irrelevant} in the magnetic theory. 
$F$-maximization has allowed us to check that the size of this window is non-zero
(this is essentially the statement that the exact R-charge curve, determined by $F$-maximization,
lies below the upper bounding red curve in Fig.\ \ref{strong}). This regime is analogous to the
regime of the free magnetic phase in four-dimensional $\NN=1$ SQCD. The standard four-dimensional
Seiberg duality is expected to work also in this phase by exchanging the 
`very-strongly' coupled electric description to the IR free magnetic 
description. In our three-dimensional example we find a similar effect. In the vicinity of the 
supersymmetry breaking point, $0<n-\lambda\ll 1$, the superpotential deformation is so strongly
coupled that it is in the verge of lifting the supersymmetric vacuum. In the dual description, where
the dual 't Hooft coupling is small, $\bar \lambda=n-\lambda \ll 1$, the superpotential deformation
is irrelevant. In this case the mismatching of symmetries, number of free decoupled operators $etc.$
suggests that the correct R-symmetry cannot be read off the deforming superpotential in the electric
description, and that one should rather use the magnetic description to study the theory in this regime
(see \cite{Kutasov:2003iy} for analogous comments in the case of the four-dimensional one-adjoint 
SQCD theory). 

It is interesting to understand how this story implements itself on the level of the free energies $F$ 
computed via localization in the regime \eqref{dualdiag}. 
A naive possibility that seems unlikely to work (for the above reasons) 
is to try to match the free energy computed via localization at $R=\frac{2}{n+1}$ in the electric
theory and the free energy computed via localization at $R$ determined by $F$-maximization 
in the $\hat {\bf A}$ theory on the magnetic side. Another possibility is that the mathematical 
identities that give rise to \eqref{dualad} are blind to the existence of a conformal window and apply
for all $N,k$ in the supersymmetric interval. In that case, outside the conformal window the matrix 
model $F$ does not refer to the actual free energy of the theory. 
Yet another possibility is that they only apply for $N,k$ 
within the conformal window, but then it is interesting mathematically to identify precisely how they 
break down outside of it. 
%The 3D-4D connection \cite{Dolan:2011rp} suggests that the second possibility is realized. 

%%%%%%%%%%%%%%%%%%%%%%%%%%%%%%%%
\section{Tests of the F-theorem at large-$N$}
\label{Ftheorem}

Recently, an analog of the $c$-theorem in two dimensions has been put forward for three-dimensional
quantum field theories in Refs.\ \cite{Casini:2011kv,Jafferis:2011zi}. 
The precise proposal, which is known as the 
F-theorem, states that $F$ (the free energy of the $S^3$ partition function we have been computing) 
always decreases along RG flows. In the absence of a general proof it is instructive to extend the tests 
in as many classes of theories as possible. In this section we test the validity of the theorem in a new set 
of RG flows that occur within the setup of the $\hat {\bf A}$ and ${\bf A}_{n+1}$ theories.

\subsection{A web of RG flows}
\label{web}

By adding relevant superpotential interactions to the one-adjoint $\hat {\bf A}$ CSM theory one 
can generate a web of RG flows. We have already discussed some of these flows. In the range
of coupling where a chiral ring operator $\tr X^{n+1}$ is relevant (there is such a range
for any $n>1$), one can use the operator to deform the $\hat {\bf A}$ theory by the corresponding
superpotential and generate a supersymmetric RG flow towards a new IR fixed point that we called 
${\bf A}_{n+1}$.

Another general class of RG flows arises by using an ${\bf A}_{n+1}$ theory as the UV fixed point
and deforming it by a general polynomial superpotential of the form
\beq
\label{webaa}
W=\sum_{i=0}^n \frac{g_{i}}{n+1-i} \tr X^{n+1-i}
~.
\eeq
The supersymmetric vacua of this theory can be found by solving the $F$-term equations
\beq
\label{webab}
W'(x)=\sum_{i=0}^n g_i x^{n-i}=g_0 \prod_{i=1}^\ell (x-a_i)^{n_i}~, ~~ \sum_{i=1}^\ell n_i=n
\eeq
where $a_i$ are parameters directly related to the coefficients $g_i$ of the superpotential polynomial.
In the far infrared the theory flows to a product of ${\bf A}_{n_i+1}$ theories specified by a partition of
the total rank $N$
\beq
\label{webac}
U(N)_k^{(n+1)} \to U(N_1)_k^{(n_1+1)} \times \cdots \times U(N_\ell)_k^{(n_\ell+1)}
~, ~~ \sum_{i=1}^\ell N_i=N
~.
\eeq
There is a supersymmetric vacuum for each of the factors in this product provided $N_i\leq n_i k$
for all $i=1,2,\ldots,\ell$.

\subsection{General predictions of the conjectured F-theorem}
\label{predict}

For each of the above RG flows the F-theorem predicts the inequality
\beq
\label{predictaa}
F_{UV}>F_{IR}
~.
\eeq
For the general flow \eqref{webac} this inequality implies
\beq
\label{predictab}
F_{N,k}^{(n+1)} > \sum_{i=1}^\ell F_{N_i,k}^{(n_i+1)}
~.
\eeq
Focusing on the planar contributions $f(\lambda;n)$ in the large-$N$ limit (see eq.\ \eqref{matchaa})
we obtain
\beq
\label{predictac}
f\left (\sum_{i=1}^\ell \lambda_i ; \sum_{i=1}^\ell n_i \right) > \sum_{i=1}^\ell x_i^2 f(\lambda_i ; n_i)
\eeq
where by definition
\beq
\label{predictad}
N_i=x_i N~, ~~ \lambda_i=x_i \lambda~, ~~ 0<x_i<1~, ~~ \sum_{i=1}^\ell x_i=1
~.
\eeq

Notice that in these flows it is in general possible to start in the UV from an ${\bf A}_{n+1}$ theory
inside its conformal window and end with a product of theories where some of them lie outside their 
conformal window. In these theories the deforming operator may be irrelevant in which case 
the true IR fixed point is the $\hat {\bf A}$ theory and the free energy should be computed accordingly.

\subsection{Tests of the F-theorem}
\label{Ftheoremtests}

We proceed to demonstrate the validity of the above inequalities in a set of examples with 
increasing complexity. We will concentrate on the large-$N$ limit where we can make use of
the numerical results of section \ref{partition}.

\subsubsection{${\hat{\bf A}}\to {\bf A}_{n+1}$, ${\bf A}_{n+1} \to {\bf A}_{n'+1}$, $n'<n$}
\label{test1}

We begin with the RG flow $\hat{\bf A} \to {\bf A}_{n+1}$. We pick $\lambda<n$ to be such that 
the deforming operator $\tr X^{n+1}$ is relevant in the UV theory. In these flows 
\beq
\label{test1aa}
R_{UV}<R_{IR}=\frac{2}{n+1}
~.
\eeq
Since $F(R_{UV})$ (determined by $F$-maximization) is a global maximum in all these cases, 
the inequality \eqref{predictaa} immediately follows.

As a simple example of the general flow \eqref{webac} we can consider the case
\beq
\label{test1ab}
U(N)_k^{n_{UV}+1} \to U(N)_k^{n_{IR}+1}
~, ~~ n_{IR}<n_{UV}~, ~~ N\leq n_{IR}k
~.
\eeq
In this case, the 't Hooft coupling $\lambda$ is constant along the RG flow. We choose it so that
it lies inside the conformal window of the UV theory, namely 
$\lambda\in [\lambda_{n_{UV}+1}^*,n_{UV}-\lambda^*_{n_{UV}+1}]$. Since 
$\lambda^*_{n_{IR}+1}<\lambda^*_{n_{UV}+1}$, it follows that the IR theory will also be within
its conformal window. As a result,
\beq
\label{test1ac}
R_{UV}=\frac{2}{n_{UV}+1}<R_{IR}=\frac{2}{n_{IR}+1}
~.
\eeq
The monotonicity of $F$ as a function of $R$ (at fixed $\lambda$) guarantees again the 
F-theorem inequality \eqref{predictaa}.

\subsubsection{${\bf A}_8\to {\bf A}_4 \otimes {\bf A}_5$}
\label{test2}

\begin{figure}[t!]
\begin{center}
\includegraphics[height=4.9cm]{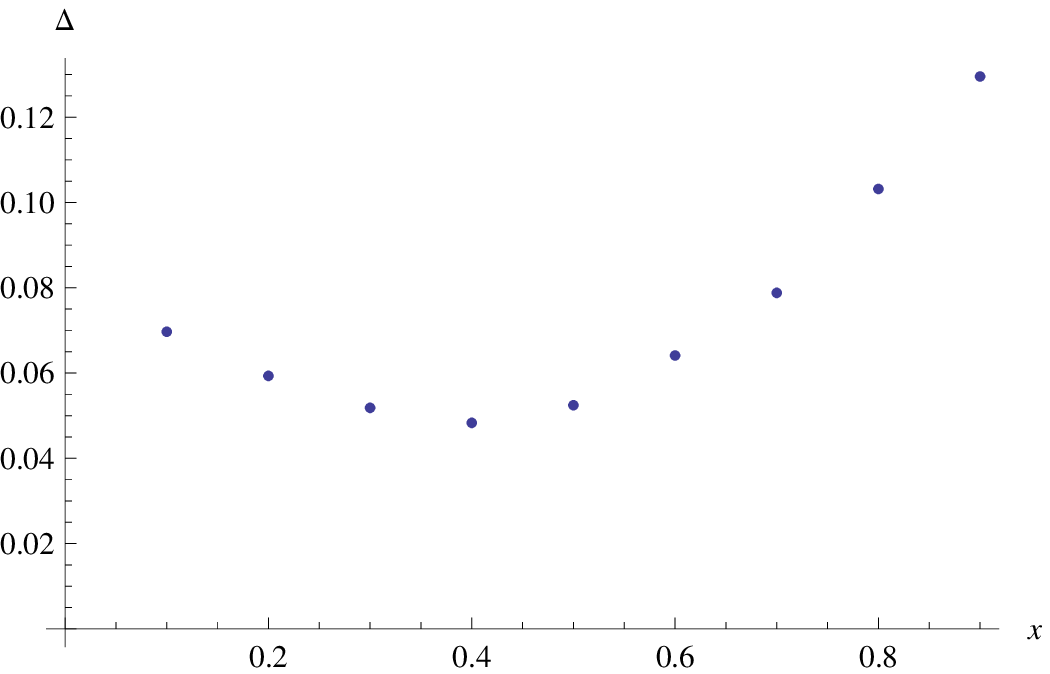}$(a)$
\includegraphics[height=4.9cm]{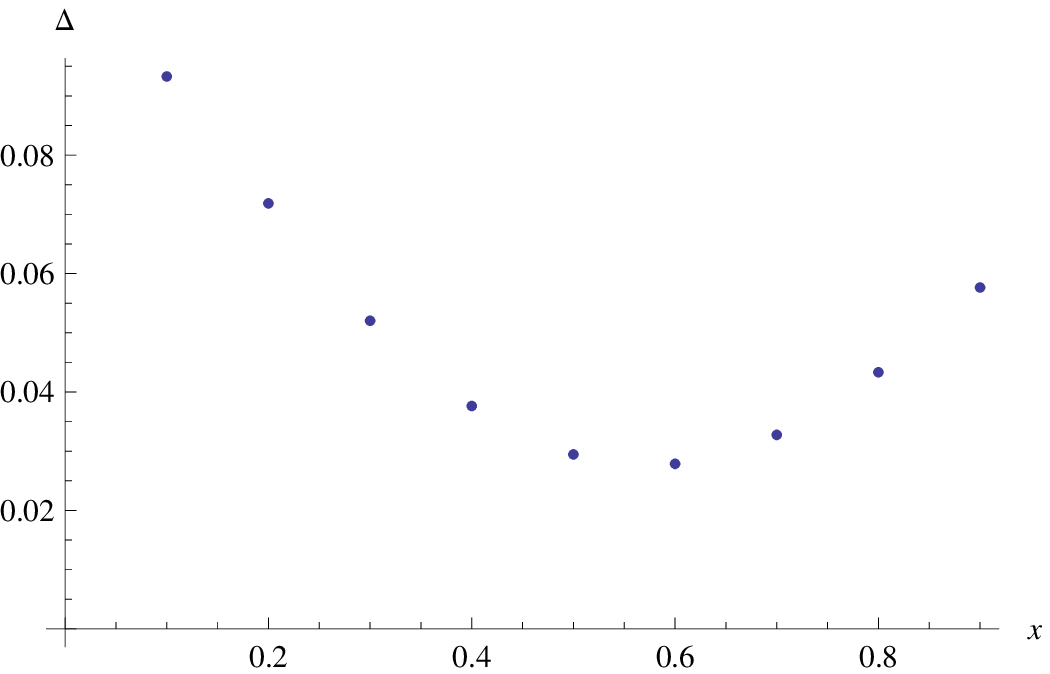}$(b)$
\end{center}
\vspace{0.3cm}
\centering{\includegraphics[height=5.5cm]{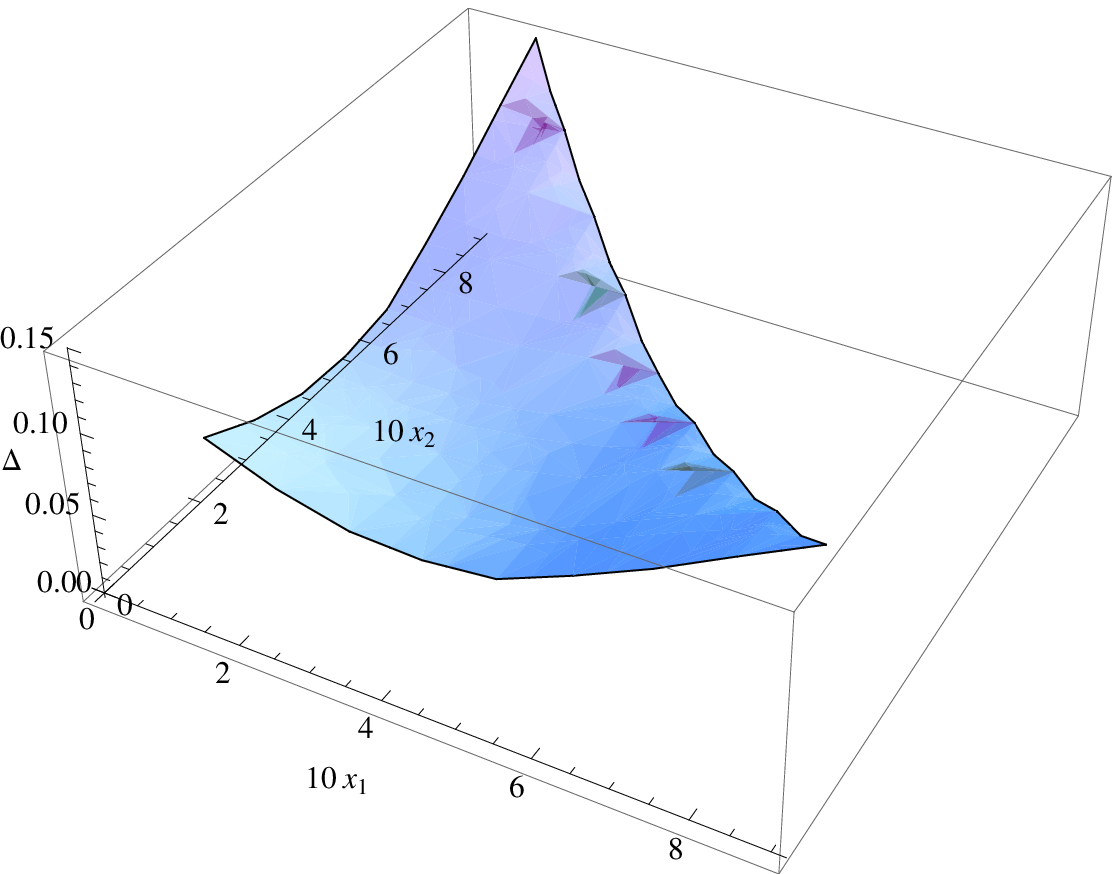} $(c)$}
\bf\caption{\it \small Plots expressing the function $\Delta$ in the case of the three RG flows
described respectively in subsections \ref{test2}, \ref{test3}, and \ref{test4}. The F-theorem requires
that $\Delta>0$, which is indeed verified. The results are based on the numerical solutions of
section \ref{partition}.}
\label{Fchecks}
\end{figure}

We move on to a more intricate one-parameter family of examples
\beq
\label{test2aa}
(n=7, \lambda) \to (n=3,\lambda x) \otimes (n=4,\lambda (1-x))
~, ~~ 0<x<1~.
\eeq
Further constraints on the parameter $x$ may arise by requiring that the IR theories have
a supersymmetric vacuum. Being obvious we will not write out these constraints explicitly
below (in the specific cases that we consider they are in any case vacuous).
The range of the conformal windows of the ${\bf A}_4,{\bf A}_5,{\bf A}_8$ theories are
\beq
\label{test2ab}
{\bf A}_4 ~:  ~(0,3) ~~~,~~
{\bf A}_5 ~:  ~ (0.35, 3.65) ~~~,~~
{\bf A}_8 ~:  ~ (1.15,5.85) 
~.
\eeq
For illustration purposes let us take $\lambda=2$ (similar results can be obtained also 
for other values of $\lambda$).
The F-theorem inequality \eqref{predictac} becomes in this case
\beq
\label{test2ac}
\Delta(x)=f(2;7)-x^2 f(2x;3)-(1-x)^2 f(2(1-x);4)>0
~.
\eeq
The results of the numerical computation are depicted in plot $(a)$ of Fig.\ \ref{Fchecks}.
We observe that the inequality \eqref{test2ac} is verified.

\subsubsection{${\bf A}_{10} \to {\bf A}_6 \otimes {\bf A}_5$}
\label{test3}

The results of a similar computation with different values of $n$ are depicted in plot $(b)$ of 
Fig.\ \ref{Fchecks}. Now we consider the one-parameter family of flows
\beq
\label{test3aa}
(n=9,\lambda) \to (n=5,\lambda x) \otimes (n=4,\lambda(1-x))~, ~~ 0<x<1
\eeq
and for concreteness we set $\lambda=3$.
The range of the conformal windows of the ${\bf A}_5,{\bf A}_6,{\bf A}_{10}$ theories are
\beq
\label{test3ab}
{\bf A}_5 ~:  ~(0,35, 3.65) ~~~,~~
{\bf A}_6 ~:  ~(0.65, 4.35) ~~~,~~
{\bf A}_{10} ~:  ~(1.7, 7.3) 
~.
\eeq

The F-theorem inequality \eqref{predictac} becomes in this case
\beq
\label{test3ac}
\Delta(x)=f(3;9)-x^2 f(3x;5)-(1-x)^2 f(3(1-x);4)>0
~.
\eeq
Again, it is apparent from the results depicted in plot $(b)$ of Fig.\ \ref{Fchecks}
that the inequality \eqref{test3ac} is verified.

\subsubsection{${\bf A}_{11} \to {\bf A}_4 \otimes {\bf A}_4 \otimes {\bf A}_5$}
\label{test4}

Finally, we report a test of the F-theorem for a two-parameter family of RG flows defined as
\beq
\label{test4aa}
(n=10, \lambda) \to (n=3, \lambda x_1) \otimes (n=3,\lambda x_2) \otimes (n=4, \lambda(1-x_1-x_2))
\eeq
with $0<x_{1,2}<1, 0<1-x_1-x_2<1$. In the case depicted in Fig.\ \ref{Fchecks}$(c)$ $\lambda=3$.
The range of the conformal windows of the ${\bf A}_4,{\bf A}_{11}$ theories are
\beq
\label{test3ac}
{\bf A}_4 ~:  ~(0, 3) ~~~,~~
{\bf A}_{11} ~:  ~(2, 8)
~.
\eeq

The F-theorem inequality \eqref{predictac} is more complicated in this case
\beq
\label{test3ac}
\Delta(x_1,x_2)=f(3;10)-x_1^2 f(3x_1;3)-x_2^2 f(3x_2;3)-(1-x_1-x_2)^2 f(3(1-x_1-x_2);4)>0
~.
\eeq
The results depicted in plot $(c)$ of Fig.\ \ref{Fchecks} verify this inequality.

\vspace{0.5cm}

One could proceed in this way to generate more tests of the F-theorem.

%%%%%%%%%%%%%%%%%%%%%
\section*{Acknowledgements}

We are grateful to Francesco Benini, Cyril Closset, Stefano Cremonesi, Nadav Drukker, 
Elias Kiritsis, David Kutasov, Marcos Mari\~no, Robert Myers, Niels Obers, Andrei Parnachev,
Jan Troost and Grigory Vartanov for helpful comments and discussions, and the organizers of the 6th 
Crete Regional meeting on String Theory in Milos, Greece, for a stimulating environment. Some parts of 
the numerical calculations in this work were carried out at  YITP in Kyoto University. VN would also 
like to thank the Centro de Ciencias de Benasque and the organizers of the workshop `Gravity - New 
perspectives from strings and higher dimensions' for hospitality during the final stages of this work. 
This work was partially supported by the European Union grants 
FP7-REGPOT-2008-1-CreteHEPCosmo-228644 and PERG07-GA-2010-268246.

%%%%%%%%%%%%%%%%%%%%%%%%%%%%%%%%%%%%%%%
\section*{Appendices}

\begin{appendix}

\section{Multi-cut saddle point configurations}
\label{multisaddle}

Our purpose in this appendix is to demonstrate that the saddle point equations \eqref{saddleaa}
have general multi-cut solutions. We will discuss the presence of these solutions in a perturbative
expansion of the equations in $\lambda \ll 1$ focusing, for illustration purposes, only on the leading 
term of the expansion. Our main concern here is not to find explicit solutions to the perturbative 
equations, but to demonstrate the allowed types of expansions. We have verified the existence of
explicit solutions numerically.

A perturbative analysis of the saddle point equations
\beq
\label{multiaa}
\II_i \equiv \frac{\tt i}{\lambda}t_i +\frac{1}{N} \sum_{j\neq i}^N 
\left[ \coth (\pi t_{ij})-\frac{(1-R)\sinh(2\pi t_{ij})+t_{ij} \sin(2\pi R)}{\cosh(2\pi t_{ij})-\cos(2\pi R)} \right]
=0~, ~~ i=1,2,\ldots, N
\eeq
together with the R-extremization equation
\beq
\label{multiab}
{\rm Re}\left[ \sum_{i,j=1}^N \left( 1-R+{\tt i} t_{ij}\right) \cot \pi (1-R+{\tt i}t_{ij}) \right]=0
\eeq
appeared in \cite{Minwalla:2011ma}. In this appendix we are interested in general values of $R$, 
hence we will not worry about the R-extremization equation.

In order to keep the discussion simple and compact we begin by listing a few characteristic expansions.

\subsection*{The standard one-cut solution crossing the origin}

This expansion is based on the ansatz
\beq
\label{multiac}
t_i=\sqrt{\lambda} \left( t_i^{(0)}+\lambda t_i^{(1)} +\cdots \right)~, ~~ i=1,2,\ldots, N
~.
\eeq
Inserting this expansion into the saddle point equations \eqref{multiaa} and expanding in powers
of $\lambda$ we get 
\beq
\label{multiad}
\II_i=\frac{1}{\sqrt \lambda} \left( {\tt i} t_i^{(0)}
+\frac{1}{\pi N} \sum_{j\neq i}^N \frac{1}{t_i^{(0)}-t_j^{(0)} }\right) +\ldots=0
~.
\eeq
The dots represent higher powers in the expansion. Ref.\ \cite{Minwalla:2011ma} explains how 
one can proceed to solve these equations systematically at any order. At leading order the equations 
(which are independent of $R$) reduce to those of the familiar Wigner model and the eigenvalues 
exhibit a semicircle distribution. 

This expansion captures the one-cut solutions that reproduce correctly the perturbative field theory
computation of the R-charge and are the key players in the numerical analysis of the main text.

\subsection*{Two-cut solutions crossing the imaginary axis at $\pm \frac{m{\tt i}}{2}$}

Consider an expansion based on the ansatz
\bea
\label{multiae}
&&t_i=\frac{m {\tt i}}{2}+\lambda t_i^{(0)}+\lambda^2 t_i^{(1)}~, ~~i=1,2,\ldots,\frac{N}{2} \in \SS^+~,
\nonumber\\
&&t_i=-\frac{m {\tt i}}{2}+\lambda t_i^{(0)}+\lambda^2 t_i^{(1)}~, 
~~i=\frac{N}{2}+1,\frac{N}{2}+2,\ldots,N \in \SS^-
~.
\eea
Inserting this expansion into the saddle point equations we find for $m\in \R_+ - \N$
\beq
\label{multiaf}
\II_{i\in \SS^{\pm}}=\frac{1}{\lambda} \left( \mp \frac{m}{2}+\frac{1}{\pi N} 
\sum_{j\neq i \in \SS^{\pm}} \frac{1}{t_i^{(0)}-t_j^{(0)}} \right)+\ldots=0
\eeq
and for $m\in \N$
\beq
\label{multiaf}
\II_{i\in \SS^{\pm}}=\frac{1}{\lambda} \left( \mp \frac{m}{2}+\frac{1}{\pi N} 
\sum_{j\neq i}^N \frac{1}{t_i^{(0)}-t_j^{(0)}} \right)+\ldots=0
~.
\eeq

One can show that only the $m\in \N$ case admits solutions. We have not attempted to find these 
equations analytically, but have verified numerically that they exist (see $e.g.$ plot $(b)$ of 
Fig.\ \ref{eigen}).

\subsection*{Two-cut solutions crossing the imaginary axis at $\pm \frac{(R+m){\tt i}}{2}$}

In this case we consider the ansatz
\bea
\label{multiag}
&&t_i=\frac{(R+m) {\tt i}}{2}+\lambda t_i^{(0)}+\lambda^2 t_i^{(1)}~, ~~i=1,2,\ldots,\frac{N}{2} \in \SS^+~,
\nonumber\\
&&t_i=-\frac{(R+m) {\tt i}}{2}+\lambda t_i^{(0)}+\lambda^2 t_i^{(1)}~, 
~~i=\frac{N}{2}+1,\frac{N}{2}+2,\ldots,N \in \SS^-
~.
\eea
Inserting this expansion into the saddle point equations we obtain to leading order
\beq
\label{multiai}
\II_{i\in \SS^{\pm}}=\frac{1}{\lambda}\left(
\mp \frac{R+m}{2} +\frac{1}{\pi N} \sum_{j\neq i \in \SS^{\pm}}\frac{1}{t_i^{(0)}-t_j^{(0)}}
-\frac{m+1}{2\pi N} \sum_{j\in \SS^{\mp}}\frac{1}{t_i^{(0)}-t_j^{(0)}}\right)+\ldots=0
~.
\eeq

\subsection*{General multi-cut expansions}

One can also entertain the possibility of more general multi-cut solutions based on the following
ansatz
\bea
\label{multiaj}
&&N_0~{\rm eigenvalues~in}~ \SS_0~~: ~~ t_i=\lambda^\alpha t_i^{(0)}+\ldots~,
\nonumber\\
&&N_m~{\rm eigenvalues~in}~ \SS_m^\pm~:~~ t_i=\pm \frac{m{\tt i}}{2}+\lambda t_i^{(0)}+\ldots~,
\\
&&M_m~{\rm eigenvalues~in}~ \SS^{\pm,R}_m~:~~ 
t_i=\pm \frac{(R+m){\tt i}}{2}+\lambda t_i^{(0)}+\ldots~,
\nonumber
\eea
so that
\beq
\label{multiak}
N_0+\sum_{m=1}^\infty 2(N_m+M_m)=N
~.
\eeq
$\alpha=\frac{1}{2},1$ are exponents with special features but other values may also be allowed.

We have not attempted to solve the above equations analytically, but have verified 
numerically the existence of multi-cut solutions of the above more general type.
In general, the actual solution of these equations will impose further constraints on the allowed 
values of the integers $N_0,N_m,M_m$.

The above list covers the full set of allowed expansions.

\section{Spontaneous SUSY breaking implies $Z_{S^3}^{\rm (loc)}=0$: outline of an argument}
\label{bargument}

In this appendix we outline a potential line of reasoning in favor of statement $(b)$ in section 
\ref{susy}. Our purpose is to show that in $\NN=2$ theories with spontaneous supersymmetry 
breaking the $Q$-deformed partition function $Z_{S^3}^{\rm (loc)}$ vanishes. We describe on
general grounds why this is a potentially correct generic property and highlight some of the key 
aspects that need to be clarified in order to reach a proof.

Let us begin with a few general comments.
For a theory on a 3-manifold $\MM$ with an $S^2$ boundary there is a natural Hilbert space $\HH$  
associated with the boundary $S^2$ \cite{Witten:1988hf}. The path integral of the theory over $\MM$ 
with boundary conditions $\Phi |_{S^2}=\chi$ 
\beq
\label{bargumentaa}
Z(\chi)=\int_{\Phi |_{S^2}=\chi} e^{-S}
\eeq
defines a vector $|\chi\rangle$ in the Hilbert space $\HH$. We are interested in the case 
where $\MM$ is a semisphere.

With the insertion of a generic operator $\OO$ at the pole of the semisphere
the path integral with boundary conditions $\chi$ gives the amplitude $\langle \OO | \chi \rangle$.
This formalism is the basis of the operator-state correspondence (see, for example, 
\cite{Ginsparg:1993is} for a review). In a bra-ket language 
\beq
\label{bargumentab}
Z(\chi)=\langle \Omega | \chi \rangle
\eeq
where $|\Omega\rangle$ is the vacuum state of the theory corresponding to the insertion of the
identity operator.

The partition function on $S^3$ can be written in the following form
\beq
\label{bargumentac}
Z_{S^3}=\langle \Omega | \Omega \rangle = \int d\chi\, Z(\chi) Z^*(\chi) = 
\int d\chi \, \langle \Omega | \chi\rangle \langle \chi | \Omega \rangle
~.
\eeq
The second equality corresponds to gluing two 3-spheres at the $S^2$ equator and summing
over the boundary conditions $\chi$. We will be applying \eqref{bargumentac} to the $Q$-deformed
path integral \eqref{susyaa}.
In this language
\beq
\label{bargumentad}
\frac{d Z_{S^3}(t)}{dt}=-\langle \Omega | \int \{ Q, V \}  | \Omega \rangle
~.
\eeq
As a result, in a theory with spontaneously broken supersymmetry $Q|\Omega\rangle \neq 0$ 
and generically the derivative $\frac{d Z_{S^3}(t)}{dt}$ is non-zero. From now on we will be referring 
exclusively to the $t=\infty$ theory whose partition function is denoted as $Z_{S^3}^{({\rm loc})}$
in the main text.

Another useful piece of information that we need is the following.
Localization is based on a supercharge $Q$. There is an additional supercharge $Q^\dagger$ with 
anticommutator
\beq
\label{algebraa}
\{ Q, Q^\dagger \}=M+R
\eeq
which is also a symmetry of the theory. $M$, which is part of the $USp(2,2)$ conformal 
group on $S^3$, is a rotation on $S^3$ that can be viewed as a translation along the Hopf fiber of 
$S^1 \hookrightarrow S^3 \to S^2$. $R$ is the R-symmetry operator.\footnote{For additional details 
we refer the reader to Ref.\ \cite{Jafferis:2010un}.}

The ground states $\Omega$ have zero R-charge. Hence, using a standard argument based on the 
correlation function 
\beq
\label{algebrab}
\langle \Omega | M | \Omega \rangle=
\langle \Omega|\{ Q, Q^\dagger \} |\Omega \rangle=\left |  Q |\Omega \rangle \right|^2+
\left |  Q^\dagger  |\Omega \rangle \right|^2
\eeq
we deduce that a ground state is supersymmetric if and only if it has $M=0$.
Notice, however, that the second equation in 
\eqref{algebrab} is strictly valid in Lorentzian signature. In Euclidean signature 
$Q^\dagger$ is not the Hermitian conjugate of $Q$, but an independent supercharge. In what
follows we will nevertheless assume that the above conclusion, $i.e.$ that a ground state is supersymmetric if and only if it has $M=0$, is also correct in the Euclidean theory. Clearly, this 
is a point that deserves further justification.

Furthermore, one can show, using an argument in \cite{Witten:1982df}, that the zero-$M$  
ground states are precisely the zero R-charge states $|\alpha\rangle$ that obey the equation
\beq
\label{statesa}
Q | \alpha\rangle =0~, ~~ |\alpha\rangle \neq Q | \beta \rangle
\eeq 
for any state $|\beta \rangle$. This concludes our short parenthesis of introductory comments.

By definition, $Z_{S^3}^{\rm (loc)}$ receives contributions from states $|\chi\rangle$
that solve the equation\footnote{More specifically, since localization forces the path integral on 
configurations $\Phi$ with the property $\{ Q,V \}\Phi=0$ the same property must be obeyed by 
continuity by the boundary conditions $\chi$.}
\beq
\label{equa}
\{Q, V \}| \chi \rangle=0 
~.
\eeq
Applying $Q$ to this equation and using $Q^2=0$ we find that the states $VQ|\chi\rangle$ 
are annihilated by $Q$, $i.e.$ that
\beq
\label{equb}
QVQ |\chi\rangle =0
~.
\eeq
Therefore, according to the above general discussion, if the states $VQ|\chi\rangle$ are 
zero R-charge states that are not $Q$-exact we may infer that they have $M=0$ and are therefore 
supersymmetric ground states. Then $(b)$ follows naturally. When supersymmetry is spontaneously 
broken there are no supersymmetric ground states. Hence, there are no zero R-charge, zero-$M$ 
ground states, including states of the form $VQ|\chi \rangle$, and therefore no states $\chi$
that obey equation \eqref{equa} that can contribute to the path integral $Z_{S^3}^{\rm (loc)}$.
As a result, $Z_{S^3}^{\rm (loc)}$ receives no contributions from the sum $\int d\chi$ in
the rightmost expression in eq.\ \eqref{bargumentac} and vanishes identically as stated in $(b)$.

The assumption that the states $VQ|\chi\rangle$ have zero R-charge is plausible for the following
reason. The implementation of localization in general Yang-Mills theories 
with arbitrary matter \cite{Kapustin:2009kz,Jafferis:2010un,Hama:2010av} and the eventual reduction
of the path integral to a matrix integral over the real, zero R-charge scalar field $\sigma$ in the $\NN=2$ 
vector multiplet, suggests that, at least in this general class of theories, $\chi$ is a state 
with zero R-charge. Since the R-charge of $VQ$ is zero\footnote{Indeed, the R-charge of the 
$Q$-deformation in \eqref{susyaa} must be zero. The explicit formulae for $V$ in 
\cite{Kapustin:2009kz,Jafferis:2010un,Hama:2010av} verify this property.} we conclude that 
$VQ|\chi\rangle$ is a zero R-charge state as well. In more general situations one has to examine 
precisely how localization works.

The other part of the argument requires showing that the state $VQ|\chi\rangle$ is not $Q$-exact,
namely that there is no state $|\beta\rangle$ for which
\beq
\label{statesb}
VQ|\chi\rangle=Q|\beta\rangle
~.
\eeq
This property seems plausible, but we have not been able to find a rigorous proof. 

It would be interesting to know if the specific choice of $V$ plays any particular role in this argument. In 
general, $Z_{S^3}^{\rm (loc)}$ is expected to depend on the choice of $V$ when a supersymmetric 
vacuum is absent, but the property $Z_{S^3}^{\rm (loc)}=0$ may not. The latter would have to be true for 
$(b)$ to hold in its current form, otherwise one would have to specify a special class of $V$ functionals 
for which $(b)$ is valid. 

Moreover, part of our proposal is that any unknown trial R-charges in 
$Z_{S^3}^{\rm (loc)}$ should be fixed by $|Z_{S^3}^{\rm (loc)}|$-minimization even in the absence of a 
supersymmetric vacuum. This prescription is motivated by the requirement to have a 
universally-prescribed quantity that reproduces the physical sphere partition function when 
supersymmetry is not broken. The implications and necessity of this prescription for the validity of $(b)$ 
should be clarified.

\section{Matching deformations in the ${\bf A}_{n+1}$ duality}
\label{defmatching}

The duality \eqref{dualaa} is expected to hold for arbitrary superpotential deformations of 
the type \eqref{webaa}. Requiring the general matching of the vacuum structure on both sides
of the duality gives useful information about the precise map between chiral ring operators.
We proceed to derive this map by suitably adapting the four-dimensional analysis of 
\cite{Kutasov:1995ss}.

We define the `electric' theory as the $U(N)_k^{(n+1)}$ theory deformed by the general superpotential
\beq
\label{defaa}
W=\sum_{i=0}^n \frac{g_{i}}{n+1-i} \tr X^{n+1-i}
~.
\eeq
The dual `magnetic' theory is a $U(nk-N)_{-k}^{(n+1)}$ theory deformed by a dual superpotential of
the form
\beq
\label{defab}
\overline{W}=\sum_{i=0}^n \frac{\bar g_{i}}{n+1-i} \tr \bar X^{n+1-i}+\alpha(\{ g_i \})
~.
\eeq
$\bar X$ denotes the chiral superfield in the adjoint representation of the dual theory. We have allowed
for a constant term in the superpotential denoted as $\alpha(\{ g_i \})$. Our purpose is to determine 
$\bar g_i$ and $\alpha$ as functions of the electric superpotential couplings $g_i$.

\subsection*{Electric theory}

Consider, for example, the case where 
\beq
\label{defac}
W'(x)=\sum_{i=0}^n g_i x^{n-i}=g_0 \prod_{i=1}^n (x-a_i)
\eeq
with all $a_i$ different.
As described in the main text, in the deep IR the theory splits into a set of decoupled ${\bf A}_2$
theories with gauge groups $U(N_i)$ labelled by the sequence of integers
\beq
\label{defad}
N_1 \leq N_2 \leq \cdots \leq N_n~, ~~ \sum_{i=1}^n N_i=N
~.
\eeq
$N_i$ eigenvalues of the matrix $X$ reside in the $i$-th minimum of the potential $V=|W'(x)|^2$
labeled by $a_i$. Each of these ${\bf A}_2$ theories has a supersymmetric vacuum if and only
if $N_i \leq k$.

\subsection*{Magnetic theory}

Similarly, on the magnetic side
\beq
\label{defad}
\overline{W}'(x)=\sum_{i=0}^n \bar g_i x^{n-i}=\bar g_0 \prod_{i=1}^n (x-\bar a_i)
\eeq
with all $\bar a_i$ different. In the deep IR the theory splits into $n$ copies of the ${\bf A}_2$
theory with gauge group $U(\bar N_i)$, such that
\beq
\label{defae}
\bar N_1 \leq \bar N_2 \leq \cdots \leq \bar N_n~, ~~ \sum_{i=1}^n \bar N_i=\bar N=nk-N
~.
\eeq

\vspace{0.5cm}

The duality must work individually for each of the $n$ components of the IR theory, hence
\beq
\label{defaf}
\bar N_i=k-N_i ~, ~~ i=1,2,\ldots, n
~.
\eeq
For this property to be realized with any choice of partitions $\{ N_i \}$ the electric and magnetic
superpotentials must be closely related. In particular, whenever any number of critical points 
$a_i$ coincides, the same number of dual critical points $\bar a_i$ must coincide as well. This
requirement puts constraints on the dual couplings $\bar g_i$.

A potential solution to these constraints is to set
\beq
\label{defag}
\bar g_i=c \, g_i ~, ~~ i=1,2,\ldots,n
\eeq
where $c$ is some constant. We adopt this solution and, following \cite{Kutasov:1995ss}, 
fix our conventions so that $c=-1$.

The duality map $g_i \to \bar g_i$ is closely related to the map between the chiral ring operators
$\tr X^i$ and $\tr \bar X^j$. More information about this map can be inferred in the following way.
Define the free energy ${\bf F}(g_i)$ of the theory in flat space as
\beq
\label{defai}
e^{-\int d^3 x d^2 \theta \, {\bf F}(g_i)+c.c.}=
\langle e^{-\int d^3 x d^2 \theta \, W(X,g_i)+c.c.}\rangle
\eeq
where $g_i$ are background chiral superfields. Then
\beq
\label{defaj}
\frac{1}{n+1-i}\langle \tr X^{n+1-i} \rangle =\frac{\d {\bf F}}{\d g_i}
~.
\eeq
Similarly, on the magnetic side
\beq
\label{defak}
\frac{1}{n+1-i}\langle \tr \bar X^{n+1-i} \rangle =\frac{\d \bar {\bf F}}{\d \bar g_i}
~.
\eeq

The duality implies 
\beq
\label{defal}
\bar {\bf F}\left (\bar g_i(g) \right)={\bf F}(g_i)
~.
\eeq
Hence, 
\beq
\label{defam}
\frac{1}{n+1-i} \tr X^{n+1-i}=\sum_{j=0}^n \frac{1}{n+1-j} \tr \bar X^{n+1-j} \frac{\d \bar g_j}{\d g_i}
+ \frac{\d \alpha}{\d g_i}
~.
\eeq
Taking the vev of both sides of this equation we find additional constraints on the map 
$g_i \to \bar g_i$. Most notably, the vevs of the left hand and right hand sides, which depend
non-trivially on the particular vacuum (namely the partition $\{N_i\}$), must satisfy a relation
independent of the particular vacuum. Implementing the ansatz \eqref{defag} (with $c=-1$)
we obtain the relation
\beq
\label{defan}
\tr X^{n+1-i}=-\tr \bar X^{n+1-i}+(n+1-i) \frac{\d \alpha}{\d g_i}~, ~~ i=1,2,\ldots, n
~.
\eeq

Taking the vev of both sides of this equation in a vacuum specified by the partition $\{ N_i \}$
and the critical points $a_i=\bar a_i$ gives (taking also the duality eqs.\ \eqref{defaf} into account)
\beq
\label{defao}
u_{n+1-i}=\frac{n+1-i}{k} \frac{\d \alpha}{\d g_i}
\eeq
where following \cite{Kutasov:1995ss} we have defined
\beq
\label{defap}
u_j \equiv \sum_{i=1}^n (a_i)^j
~.
\eeq
Solving \eqref{defao} one finds
\beq
\label{defaq}
\alpha=\frac{k}{n+1} \sum_{i=1}^n \left( i g_i \frac{u_{n+1-i}}{n+1-i}\right)
~.
\eeq
The following identities are useful in deriving this result
\beq
\label{defar}
\frac{\d}{\d g_i} \left( \frac{u_{i+m}}{i+m} \right)={\rm independent~of~}i~, ~~
\sum_{i=1}^n \left ( i g_i \frac{\d u_j}{\d g_i} \right)=j u_j
~.
\eeq

We conclude that the operator map takes the form
\beq
\label{defas}
\tr X^{n+1-i}=-\tr \bar X^{n+1-i}+k u_{n+1-i}~, ~~ i=0,1,\ldots, n
\eeq
and the superpotentials $W$, $\overline{W}$ match with the identifications \eqref{defag}
and \eqref{defaq}. In the special case where $g_i=0$ $(i=1,\ldots,n)$, $g_0\neq 0$ the map
\eqref{defas} becomes simply
\beq
\label{defat}
\tr X^{n+1-i}=-\tr \bar X^{n+1-i}~, ~~ i=0,1,\ldots, n
~.
\eeq

\end{appendix}

%\newpage
%%%%%%%%%%%%%%%%%%%%%%%%%%%%%%

\end{document}